# Small-scale structure of dark matter and microlensing

A V Gurevich, K P Zybin, V A Sirota

## Contents

**Abstract.** It has been revealed using microlensing that a considerable part, possibly more than half, of the dark matter in the halo of our Galaxy consists of objects with a mass spectrum ranging from 0.05 to 0.8 of the solar mass. What is the nature of these objects? There exist two hypotheses. According to one, these are Jupiter type planets or small stars (brown and white dwarfs) consisting of normal baryonic matter. According to the other, these are non-compact objects, i.e., small-scale formations in non-baryonic dark matter. Here, a theory is proposed describing the possibility of the existence of non-compact objects in the halo of our Galaxy, their structure and formation from non-baryonic matter. The theory of microlensing on compact and non-compact objects is considered in detail. The results of microlensing observations are described and compared with theory. Possible astrophysical manifestations of the presence of small-scale structure are pointed out. The field is being extensively studied and is of fundamental interest for cosmology and astrophysics.

**A V Gurevich, K P Zybin, V A Sirota**  P N Lebedev Physics Institute,
Russian Academy of Sciences,
Leninskiĭ prosp. 53, 117924 Moscow, Russia
Tel. (7-095) 132 64 14, 132 60 50, 132 61 71
E-mail: alex@td.lpi.ac.ru, zybin@td.lpi.ac.ru,
         sirota@alex.lpi.ac.ru

## 1. Introduction

Observations show that the greater part (over 90%) of the matter in the Universe is made up of invisible (dark) matter [1], and it remains unknown what particles it consists of. From analysis of the processes of primary nucleosynthesis and galaxy formation it follows, however, that dark-matter particles must be of non-baryonic origin and must interact rather weakly both among themselves and with baryonic matter. These particles are typically assumed to be either the hypothetical heavy particles predicted by supersymmetry theory (customarily referred to as neutralinos), or light axions, or cosmic strings. They all may constitute the so-called cold dark matter (CDM) [2, 3].

Non-dissipative dark matter plays a decisive role in the formation of the large-scale structure of the Universe, i.e., galaxies, clusters of galaxies, and superclusters. At the nonlinear stage of development of initially small perturbations there occurs a gravitational compression and a subsequent kinetic mixing of the dark matter leading to the formation of stationary self-trapped objects with a singular density distribution $\rho$ at the center [4, 5]

$$\rho(r) = Kr^{-\alpha}, \quad \alpha \approx 1.8. \tag{1}$$

Here $K$ is a parameter depending on the instant of time when the object was formed. It is of importance that the form of initial perturbation does not practically affect the scaling parameter $\alpha$.

The dissipative baryonic matter, which makes up but a small fraction of the total mass of the matter, is located in the center, i.e., in the peak of density, where it forms galaxies, while the non-dissipative dark matter, distributed according to the law (1), forms a giant galactic halo. This theory is confirmed by the observational data [6–8].

Owing to gravitational forces, galaxies unite to form clusters which, in turn, form large Abell clusters and superclusters thus showing up hierarchical clumping. The dominant role in this process is also played by dark matter in which, irrespective of the object's size, the distribution (1) inevitably sets in. This property of hierarchical clumping is clearly pronounced in the pair correlation function of galaxies and other objects. At this point, the theory [9] also agrees well with the observational data [1, 10, 11].

We have up to now considered clumping of large-scale objects. What happens to objects that are smaller than galaxies? How far can the hierarchical structure extend towards small scales? It had normally been assumed that the minimal objects formed of dark matter were halos of small galaxies with a mass of $(10^7 - 10^8) M_\odot$. However, the ideas of possible scales of minimal objects have lately changed radically. In the papers by the present authors [12, 13] it is shown that dark matter can form very small gravitationally bound objects. The mass of these objects is determined by the structure of the spectrum of initial fluctuations. The structure of the spectrum on small scales is now unavailable. On the basis of the hypothesis [12, 13] concerning the character of the initial spectrum in this region, it was demonstrated that in the dark matter on these scales a hierarchical structure fairly similar to the large-scale structure may develop, extending to quite a large amount of objects with masses as large as that of the Sun.

The question, however, is whether such dark-matter objects can exist now, and if they can, in what form. How long do they live and how are they distributed in the halo of the Galaxy? The analysis of these questions shows [13] that the lifetime of small-scale objects is much longer than the age of the Universe. They are distributed in the halo according to the law (1). Thus, the theory points out the possibility of the existence of a large amount of small-scale non-baryonic objects in the Galactic halo with a fairly broad mass spectrum.

Manifestations of dark matter on large scales up to now have been registered only in observations of the dynamics and gravitational lensing of large-scale structures. Experiments on microlensing which revealed that a substantial amount of invisible objects of mass $M \sim (0.05-0.8) M_\odot$ [14–16] exist in the halo of our Galaxy are therefore a fundamental new step. These are usually supposed to be brown or white dwarfs or planets ('Jupiters') consisting of ordinary baryonic matter [15, 17].

However, the number of dark objects thus revealed appeared to be very large. A detailed analysis carried out in recent papers (see Ref. [16]) demonstrates that the observed objects make up perhaps over 50% of the total mass of dark matter in the halo of our Galaxy. So, we find a contradiction with the conventional ideas of the non-baryonic nature of dark matter. Furthermore, with such an interpretation one encounters some difficulties in what concerns the consistency between microlensing data and direct optical observations from the Hubble telescope [18].

In this connection it was hypothesized in Ref. [12] that microlensing in the halo reveals not Jupiter type objects and cold stars, but objects of small-scale hierarchical dark-matter structures consisting of non-baryonic matter. These objects, however, are non-compact and have sizes of the order of the Einstein radius, characterizing microlensing, or may even be several times greater. For this reason, the theory of microlensing by non-compact dark-matter objects was developed in Ref. [19] (see also Ref. [20]). A detailed comparison of this theory with the available observational data gives a good agreement [19, 13]. It should nevertheless be emphasized that the same observational data are in agreement with the theory of microlensing by compact objects. The difference in the results of these theories now lies within the experimental error. Therefore, the question of the presence of both non-compact and compact objects in the halo of our Galaxy remains open.

Our aim here is to give a review of the present state of this problem. We should stress that if in comparison with the theory some non-compact objects are discovered as a result of improved precision and processing of microlensing observation, this will simultaneously mean a direct discovery of non-baryonic CDM for the reason that no baryonic objects of such a mass, scale and luminosity may exist.

## 2. Minimal dark-matter objects

### 2.1 Primary spectrum cut-off
Dark-matter objects are formed upon development of Jeans instability: as a result of gravitational attraction, small initial perturbations of a uniformly distributed matter increase. Crucial for the formation of such a structure is the form of the initial density fluctuations

$$\delta(\mathbf{x}, t) = \frac{\rho(\mathbf{x}, t) - \rho_0(t)}{\rho_0(t)}, \tag{2}$$

where $\rho(\mathbf{x}, t)$ is the local density of matter and $\rho_0(t)$ is the mean density of matter. The initial fluctuations $\delta_{i0}(\mathbf{x})$ are customarily represented by the Fourier spectrum $|\delta_i(\mathbf{k})|^2$, where

$$\delta_i(\mathbf{k}) = \int_{-\infty}^{+\infty} \delta_{i0}(\mathbf{x}) \exp(i\mathbf{k}\mathbf{x}) \, d\mathbf{x}. \tag{3}$$

The form of the spectrum of initial fluctuations is determined by physical processes in the early Universe. The spectrum

$$|\delta_i(\mathbf{k})|^2 \propto k \tag{4}$$

is used conventionally, suggested in 1965 by Ya B Zel'dovich [21] and confirmed by recent observations [22] over large scales.

The spectrum (4) is cut off for low values of the wave number $k$ on the scale of the horizon radius $R_h^{-1}$. For high values of the wave number $k \to \infty$, the cut-off occurs during decoupling, i.e., in the period when the particles of the dark matter escape thermal equilibrium. Note that one typically considers both HDM and CDM (or a combination). HDM is understood as weakly interacting particles that leave thermodynamic equilibrium when they are still relativistic (for example, light neutrinos). CDM is thought of as consisting of rather heavy particles of mass $m_x \gg 10^2$ eV whose strong interaction stops as soon as the temperature in the expanding Universe lowers to $T \approx m_x c^2$. The concentration of such particles may continue to decrease for some time at the expense of annihilation. In both cases, the spectrum varies on small scales during freezing-out of dark-matter particles. A free kinetic spread of these particles results in a smearing of the fluctuations whose scale is less than the horizon radius at the moment $t_x$, when the temperature is

$$T(t_x) = m_x c^2. \tag{5}$$

It is the horizon radius $R_h^{t_x}$ at this instant of time that determines the characteristic scale of the cut-off [23]:

$$k_{\max} \approx \frac{1}{R_h^{t_x}}. \tag{6}$$

On large scales $k \ll k_{\max}$, the fluctuation spectrum (3), (4) remains practically unaltered, whereas on small scales $k > k_{\max}$ the free spread of particles causes a strong damping of fluctuations, which leads to a sharp fall of the spectrum on the scales $l < k_{\max}^{-1}$. The calculated variations of the spectrum (4) due to this process are presented in Fig. 1 [23]. The position of the peak is determined by relations (5) and (6). The peak is fairly broad:

$$0.5 k_{\max} < k < 2 k_{\max}. \tag{7}$$

After this peak, a sharp fall of the spectrum is observed. The maximum in the spectrum and the steep fall when $k > k_{\max}$ are indicative of the fact that in the realization of the spectrum there are no perturbations with scales below $l_{\min} \approx (2 k_{\max})^{-1}$. Consequently, the smallest objects formed of dark matter will just have scale $l_{\min}$.

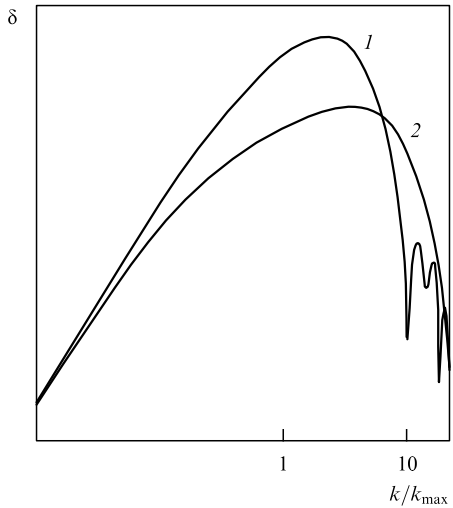

**Figure 1.** Zel'dovich–Harrison spectrum transformation due to free flight of dark-matter particles [23]: (*1*) statistical weight $g = 30$, (*2*) $g = 200$ (16).

## 2.2 Linear growth of fluctuations

The small-scale dark-matter structure, as for the large-scale, appears owing to the growth of small fluctuations upon the development of Jeans instability. An important peculiarity of this process is that at the radiation-dominated stage of Universe expansion, the increase of fluctuations is slow. Their comparatively rapid growth starts just after equilibrium has set in and the expansion is already determined by matter rather than radiation [24].

We shall discuss the dynamics of the growth of dark-matter fluctuations. For simplicity we shall assume the parameter $\Omega$ to be equal to unity, that is, the density of matter to be equal to the critical density. In this case, on scales less than the horizon, one can use the Newtonian approximation in the analysis of fluctuation dynamics. The equations describing the linear growth of density fluctuations $\delta(\mathbf{x}, t)$ take a simple form [1]

$$\frac{d^2 \delta}{dt^2} + \frac{2}{a} \frac{da}{dt} \frac{d\delta}{dt} = \frac{3}{8} \frac{\epsilon}{a^3} \delta, \quad \left(\frac{1}{a} \frac{da}{dt}\right)^2 = \frac{1}{4}\left(\frac{1}{a^4} + \frac{\epsilon}{a^3}\right). \tag{8}$$

Here $a(t)$ is the scale factor determining the expansion of the Universe. The parameter $\epsilon$ describes the matter-to-radiation density ratio $\rho_x / \rho_\gamma$. Before the moment $t_{\rm eq}$ of equilibrium onset, when $a < a_{\rm eq}$ and

$$a(t_{\rm eq}) = a_{\rm eq} = \frac{1}{\epsilon}, \quad \epsilon = \frac{\rho_x}{\rho_\gamma}, \tag{9}$$

the dominating particles are photons [the first summand in Eqn (8) for $a(t)$] and after the equilibrium onset, i.e. for $t > t_{\rm eq}$, nonrelativistic X-particles are predominant. At the radiation-dominated stage, that is, in the period when radiation is predominant, the scale factor $a(t)$ increases as $\sqrt{t}$, while at the dust stage, that is, in the period when matter is predominant, it increases as $t^{2/3}$ [1]. Since the density of matter is $\rho_x \propto a^{-3}$ and the radiation density is $\rho_\gamma \propto a^{-4}$, then $\rho_x / \rho_\gamma \propto a \propto (1+z)^{-1}$. As the present-day value of $\rho_x / \rho_\gamma$ is known, it is easy to find the density of matter and the red shift at the instant of time $t_{\rm eq}$:

$$\rho_x \approx \rho_\gamma \approx \rho_{\rm eq} \approx 3 \times 10^{-16} \text{ g cm}^{-3},$$
$$t_{\rm eq} = t(z_{\rm eq}), \quad z_{\rm eq} \simeq 3 \times 10^4. \tag{10}$$

In equations (8), the normalization of the scale factor $a(t)$ is chosen so that we have $a = 1$ at the instant of time $t_x$ when dark-matter particles escape thermal equilibrium [this instant of time is defined by Eqn (5)]. Accordingly $\epsilon = \rho_x^0 / \rho_\gamma^0$, where $\rho_x^0$ and $\rho_\gamma^0$ are the respective densities of X-particles and photons at the instant $t_x$. Since the temperature is $T \sim 10^2$ eV at the moment of equilibrium $t_{\rm eq}$, it follows that if the particle mass is $m_x \gg 10^2$ eV, then $t_x \ll t_{\rm eq}$. This means that in CDM the moment $t_x$ is attributed to the radiation-dominated stage and $\epsilon \ll 1$.

In equation (8) for $\delta$ one can get rid of the time by introducing the variable

$$\mu = \sqrt{1 + a \epsilon}.$$

Indeed, expressing $dt$ from the second equation of the system (8) in terms of $da$, we derive a universal equation for $\delta$:

$$\frac{d}{d\mu}\left[(1 - \mu^2) \frac{d\delta}{d\mu}\right] + 6\delta = 0. \tag{11}$$

The solution to this equation is the second Legendre polynomial:

$$\frac{\delta}{\delta_0} = 3\mu^2 - 1 = 3\epsilon a(t) + 2. \quad (12)$$

It should be emphasized that $\delta$ is normalized here to a small quantity $\delta_0$ which is the initial value of fluctuations at the moment $t_x$. Expression (12) describes the linear growth of fluctuations both at the radiation-dominated and the dust stages. We see that, during the radiation dominated stage, the fluctuations grow slowly, increasing only by 2.5 times up until the moment $z = z_{eq}$, i.e. $a = a_{eq}$. After this, a rapid growth begins

$$\frac{\delta}{\delta_0} \propto 3\epsilon a \propto t^{2/3}. \quad (13)$$

### 2.3 Mass of the minimal dark-matter objects

The linear growth of the fluctuations stops as soon as the magnitude of perturbations becomes comparable with the mean density of matter $\delta \sim 1$. After this, the stage of gravitational compression and kinetic mixing sets in and leads to the formation of stable spherically symmetric objects in the center of which a singular density distribution (1) is observed. This process was considered in detail in the previous review by the present authors [5]. As a result, a hierarchical structure of gravitationally bounded objects develops from the primary spectrum of fluctuations.

It is of importance that this structure is limited from below by the mass $M_{min}$ of minimal objects:

$$M_{min} \approx \rho_x(t_x) a_x^3. \quad (14)$$

Here $\rho_x(t_x)$ is the dark-matter density at the moment of decoupling, when the particles $m_x$ (5) are singled out, and $a_x$ is the scale of the horizon at the same instant of time, $a_x = R_h^{t_x}$. The instant $t_x$ in Eqn (14) depends on the mass of dark-matter particles. Consequently, it is the mass $m_x$ that exerts a crucial influence on $M_{min}$. In the case of HDM, when $m_x \ll 10^2$ eV, we have, as shown in Ref. [23],

$$M_{min} \approx \frac{m_{pl}^3}{m_x^2}. \quad (15)$$

Here $m_{pl} = (\hbar c/G)^{1/2}$ is the Planck mass. One can see that in the case under consideration the mass $M_{min}$ is related to the mass of dark-matter particles in a one-to-one manner and is rather large for HDM. For example, for the hypothetical small-mass neutrinos with $m_x = 10$ eV (which were considered in Ref. [2]), the mass is $M_x \approx 10^{17} M_\odot$, which corresponds to the mass of large clusters.

For the case of CDM for $m_x \gg 10^2$ eV, relation (15) changes (see Refs [25, 13]) to become

$$M_{min} = p \frac{m_{pl}^3}{m_x^2}, \quad p = \frac{n_x}{n_\gamma} \left(\frac{g}{4}\right)^{2/3}. \quad (16)$$

Here $n_x$ and $n_\gamma$ are the respective concentrations of CDM particles and photons, $g$ is a statistical weight of the order of $2 \times 10^2$. The factor $p$ in Eqn (16) is due to annihilation of a majority of the matter in the period after decoupling. It can be estimated from the present-day relic photon-to-baryon density ratio $(\rho_\gamma/\rho_b \approx 10^{-9})$:

$$p \approx 10^{-8} \frac{m_b}{m_x},$$

where $m_b$ and $\rho_b$ are the mass and the density of baryons.

This implies that if dark matter consists of heavy particles $m_x \gtrsim 1$ GeV, the mass of the minimal objects is very small:

$$M_x \sim (10^{-6} - 10^{-7}) M_\odot \left(\frac{m_x}{1 \text{ GeV}}\right)^{-3}. \quad (17)$$

The estimate (6), (17) gives the spectrum cut-off scale and the mass of a possible minimal object for the assumption that both the thermodynamic dark-matter particle decoupling and the cessation of interaction of these particles with others takes place at the time instant $t_x$ (5). If the decoupling and especially interaction cessation take up more time (which is the case, for example, for supersymmetric particles — neutralinos), the cut-off scale and the mass of the minimal object may increase substantially.

Another possibility for the appearance of small-mass non-baryonic structures is due to axions. The nonlinear effects of evolution of the axion field in the early Universe which may lead to the formation of gravitationally bound mini-clusters were investigated in Ref. [26]. These mini-clusters have a characteristic mass $M \sim 10^{-12} M_\odot$ and size $R \sim 10^{10}$ cm.

### 2.4 Scale of the minimal objects

As we have seen above, the mass of the minimal objects does not depend on the spectrum of initial perturbations but is determined only by the dark-matter particle mass. The scale of the minimal objects is determined, on the contrary, by the amplitude of the fluctuation spectrum $\delta$. Indeed, the mean density of matter $\rho_0(t)$ in the Universe decreases with time. Because the mass of an object is fixed [see Eqns (15), (16)], its size is naturally determined by the dark-matter density at the instant of time $t_c$ when it was formed. The quantity $t_c$ is specified by the condition of transition to the nonlinear stage of compression in the region of the spectrum maximum $\delta_m$ where the minimal objects are generated:

$$\delta_m(t_c) \sim 1. \quad (18)$$

Thus, the scale $R_x$ of a minimal object is defined as

$$R_{min} \approx \left[\frac{M_{min}}{\rho_0(t_c)}\right]^{1/3}, \quad (19)$$

where $\rho_0(t_c)$ is the dark-matter density at the moment of its formation $t_c$.

Within the period when radiation prevails, fluctuations increase very slowly (see Section 2.2). Since the initial fluctuations $\delta_0$ are not large, relation (18) can be fulfilled only provided that $t_c > t_{eq}$, where $t_{eq}$ is the moment of transition from the radiation-dominated to the dust stage, which is determined by condition (9). The red shift $z_c$ that corresponds to the moment $t_c$ is defined here by relations (12) and (18). They imply that

$$z_c + 1 = \frac{3}{2} \frac{z_{eq} + 1}{1/\delta_0 - 1}, \quad \delta_0 < 1. \quad (20)$$

Since the density in the Universe decreases rapidly as

$$\rho = \rho_{eq} \left(\frac{1+z}{1+z_{eq}}\right)^3, \quad (21)$$

where $\rho_{eq} \approx 3 \times 10^{-16}$ g cm$^{-3}$ is the density of matter at the moment of equilibrium, it follows that the scale of the minimal objects increases actively as $\delta_0$ decreases:

$$R_m = \varkappa_2 \left\{ \frac{M_{\min}}{4\pi/[(3-\alpha)\rho_{eq}]} \right\}^{1/3} \frac{2}{3}\left(\frac{1}{\delta_0} - 1\right). \tag{22}$$

This expression is specified for the scale $R_m$, namely, it now allows for the factor $\varkappa_2 \approx 0.1-0.3$ of nonlinear compression in the course of the object's formation (see Section 4.1) and the density distribution (1) inside the object.

For example, for $\delta_0 \approx 10^{-3}$ the corresponding red shift at which the object was formed (21) is $z_c \approx 1.5 \times 10^{-3} z_{eq}$, and the scale of the object of mass $M_m = 10^{-6} M_\odot$ is $R_m \approx 5 \times 10^{15}$ cm. If $\delta_0 \approx 0.4$, then $z_c \approx z_{eq}$ and the scale of the same object is much smaller: $R_m \approx 10^{13}$ cm. We also note that the scale of objects with a characteristic mass $M_m = (0.1-0.5)M_\odot$, in which we are interested later, is $(0.6-3) \times 10^{18}$ cm when $\delta_0 = 10^{-3}$ and

$$R_m \approx (3-15) \times 10^{14} \text{ cm} \tag{23}$$

when $\delta_0 = 0.4$.

## 3. Hierarchical structure

Up to now we have only discussed minimal dark-matter objects. They are first to appear. The presence of a wide spectrum of perturbations leads to the subsequent generation of objects on various scales. Large-scale objects can trap smaller-scale ones and be themselves trapped by still larger objects. Exactly this type of hierarchical structure is observed in large-scale formations in the Universe, such as galaxies, clusters of galaxies, large Abell clusters, superclusters, and so on. It is quite natural to expect that an analogous hierarchical structure occurs on small scales as well.

The sequence of objects formation on different scales is determined by the fluctuation spectrum. Going over from the Fourier transform and back to **x**-space corresponds to integration over $d^3k$ and is, therefore, equivalent to multiplication of the fluctuation spectrum by $k^3$. So, instead of $\delta(k)$ [see Eqn (4)] it is reasonable to consider the spectral function

$$F_0(k) = k^3 |\delta_i(k)|^2. \tag{24}$$

An important transformation of the fluctuation spectrum is connected with peculiarities of CDM. The point is that in the case of CDM the moment of particle decoupling and the moment when the dark-matter density begins to prevail differ appreciably in energies (and, accordingly, in time or red shifts $z$). As shown above [see Eqns (8), (12)], at the radiation-dominated stage in dark matter inside the horizon $R_h$, fluctuations increase very slowly; they are almost 'frozen'. Outside the horizon, i.e. for $k \leqslant k_{eq} = 1/R_h$, where fluctuations are small, they increase gradually in proportion with time [1]:

$$\frac{\delta\rho}{\rho} \propto t.$$

Owing to this process, fluctuations on small scales are leveled, and the spectral function is transformed as follows [1, 2]:

$$F(k) = \frac{F_0(k)}{1+(k/k_{eq})^4}. \tag{25}$$

It is the fluctuation $F(k)$ that determines sequence of appearance of objects on different scales ($l \sim 1/k$) in the hierarchical CDM structure: the first to appear are objects with maximum $F(k)$ values, and subsequently appear other, larger objects.

The form of the function $F(k)$ for the Zel'dovich–Harrison spectrum (4) is shown in Fig. 2 (curve 1). As is clear from Eqns (24) and (25), for $k > k_{eq}$ it gradually transforms to a plain form. This picture is on the whole confirmed by contemporary observational data over large scales ($l \sim 500-1000$ Mpc) [22]. Recent temperature measurements in the region of the fluctuation maximum, that is, on scales $l \sim 30-50$ Mpc, as well as on smaller scales are indicative of possible deviations from the Zel'dovich–Harrison spectrum, which in the simplest form are reduced to a variation of the exponent of the spectrum [27]:

$$|\delta_i(k)|^2 \propto k^n, \quad n \approx 1.2-1.4. \tag{26}$$

Such a power-law spectrum is also presented in Fig. 2 (curve 2). Other notable deviations of the spectrum of initial fluctuations from the Zel'dovich–Harrison spectrum are revealed in observations of distribution of the density of matter on scales of 30–50 Mpc [3].

It is, however, of importance that between the regions where the spectrum has already been measured $l \geqslant 10^{25} - 10^{26}$ cm and the region of minimum-scale objects $M \sim (10^{-6} - 1)M_\odot$, that is, $l \sim 10^{13} - 10^{15}$ cm, which is of particular interest for us, there is a huge region with a spectrum whose state is practically unknown. Note that the function $F(k)$ is strictly flat in the region $k > k_{eq}$ for the Zel'dovich–Harrison spectrum only. It may be assumed, for example, that for $k$ which are sufficiently large but below the cut-off region $k \ll k_{\max}$, the spectral function $F(k)$ has a new region where it ascends. Such a non-standard spectrum is shown in Fig. 2 (curve 3).

In the first model, curve 1 (let us call it standard), a very broad spectrum of objects, from minimal bodies to galaxies, appear almost simultaneously in the hierarchical structure. In the second model (which we refer to as power-law), first to develop are the small-scale objects, and subsequently appear structures of much larger scales. This difference in the character of object formation on different scales becomes even greater in the third (non-standard) model where small-scale structures develop at a moment close to $t_{eq}$ and large-scale structures develop much later. It should be stressed that the indicated differences in the course of object formation may have an appreciable effect on the conditions of their further existence, which determines to a large degree the form of the small-scale structure nowadays.

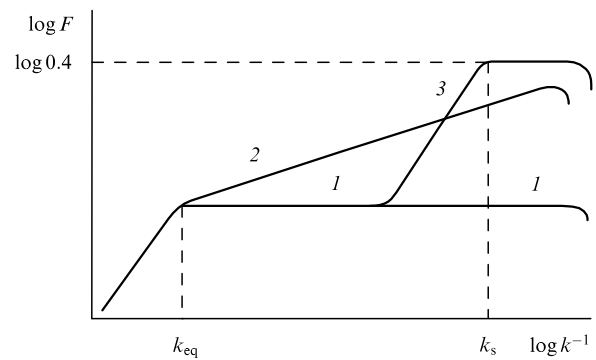

**Figure 2.** Form of the spectral function $F(k)$: (1) for the Zel'dovich–Harrison spectrum, (2) for a power-law spectrum, (3) for a non-standard spectrum.

### 3.1 Clustering

The probability that a small-scale object will be trapped by a large-scale one, provided that the difference in size is not very big, does not generally prove to be high. In the case of a power-law spectrum, this probability is determined by the clustering parameter $\varepsilon$, that is, the probability that an object of scale $R_x$ appears inside another object of size $R_f$. An estimate of the clustering parameter $\varepsilon$ has the form [5]:

$$\varepsilon = 4\pi\beta \ln \frac{R_f}{R_x}, \quad \text{where} \quad \beta < 0.016 \left(\frac{6}{m+5}\right)^{3/2}. \quad (27)$$

Here $m = n - 4$ is the exponent of the spectral function. This deviation of the exponent of the spectrum from the value (26) is due to the spectral function transformation (25). Expression (27) gives a theoretical estimate 'from above' for the clustering parameter [28]. The estimation of the parameter $\beta$ obtained from comparison with observations for a large-scale structure yields [29]

$$\beta = \beta_0 \left(\frac{6}{m+5}\right)^{3/2}, \quad \beta_0 \approx (3-6) \times 10^{-3}. \quad (28)$$

A successive entrapment of smaller objects by objects of increasingly large scales presents a picture of hierarchical clumping. The relation between the size and the mass in the case of hierarchical clumping is given by [12]

$$\frac{R_x}{R_f} = \left(\frac{M_x}{M_f}\right)^{(m+5)/6}. \quad (29)$$

Proceeding from relations (27)–(29) one can determine the mass ratio of larger $M_f$ and smaller $M_i$ objects to fit a probability of the order of unity that the larger object will entrap the smaller one. Assuming $\varepsilon \approx 1$ in formula (27) and allowing for Eqns (28) and (29), we find

$$\frac{M_f}{M_i} = \exp\left[\frac{(m+5)^{1/2}}{4\pi 6^{1/2}\beta_0}\right]. \quad (30)$$

So, for the Zel'dovich–Harrison spectrum ($n = 1$) we have

$$\frac{M_f}{M_i} = 1.9 \times 10^3 \quad \text{for} \quad \beta_0 = 6 \times 10^{-3},$$

$$\frac{M_f}{M_i} = 3.6 \times 10^6 \quad \text{for} \quad \beta_0 = 3 \times 10^{-3} \quad (31)$$

and for a sharply increasing spectrum ($n = 5$)

$$\frac{M_f}{M_i} = 3.7 \times 10^5 \quad \text{for} \quad \beta_0 = 6 \times 10^{-3},$$

$$\frac{M_f}{M_i} = 1.4 \times 10^{11} \quad \text{for} \quad \beta_0 = 3 \times 10^{-3}.$$

According to the theory, all the entrapped objects will be distributed inside the large object according to the law (1):

$$n(r) = \frac{3-\alpha}{4\pi R_f^3} N_f \left(\frac{r}{R_f}\right)^{-\alpha}. \quad (32)$$

The total number of objects $N_f$ can be evaluated from the relation

$$N_f = f \frac{M_f}{M_i}, \quad (33)$$

where $f$ is the fraction composed of objects $M_i$ in the total mass of the object $M_f$.

### 3.2 The lifetime of entrapped objects

We note first of all that objects of a hierarchical structure undergo almost no changes from the moment when they are formed till the moment when they are trapped by larger objects. Indeed, during this period they have very small peculiar velocities because the relative velocities at the moment of formation are practically zero and the distance between the objects increases owing to the general cosmological expansion. The relative velocity and the possibility of collision of objects are exclusively due to their entrapment in the gravitational field of larger-scale formations. So, we may speak only of the lifetime of objects trapped in larger-scale structures.

Entrapped objects may experience interactions. Elastic collisions between objects proceeding without energy exchange may affect only the distribution (1) of matter in them, whereas it is inelastic processes that are responsible for destruction. The principal mechanism of destruction is tidal interaction, occurring when objects fly by one another or by stars at close distances. For the case of ordinary stars, where energy dissipation takes place inside a star, the role of tidal interaction is described in detail in Ref. [30]. In our case, too, the physics of the destruction process is related primarily to tidal forces because these forces are responsible for the increase of kinetic energy of particles inside objects, and if this energy becomes equal to the binding energy of the object, we observe a destruction. Let us investigate this process more closely.

Suppose a non-compact object of mass $M_x$ and scale $R_x$ interacts with another object flying by at the distance $R$ or a massive body (star) of mass $M$. The velocity $V$ of their relative motion greatly exceeds the particle velocity $v_x$ inside the indicated non-compact object $[v_x \sim (GM_x/R_x)^{1/2}]$. Therefore, in the first approximation the particles can be considered to be at rest.

As a result of the interaction, the object $M_x$ as a whole gains additional velocity, which means an elastic scattering of objects, and moreover the particles inside it gain the additional velocity of relative motion. This is the tidal interaction. The force of tidal interaction $F_t$ is connected with the finite dimension of the non-compact body and is determined by the difference of forces acting on its particles:

$$F_t = \frac{2GMM_x}{R^3} R_x.$$

The shift of the particles under the action of this force is

$$\Delta x = \frac{F_t}{2M_x}(\Delta t)^2 = \frac{GM}{R^3} R_x (\Delta t)^2, \quad \Delta t = \frac{R}{V}.$$

Accordingly, the energy dissipated in one event is

$$\Delta E = F_t \Delta x = \frac{2G^2 M^2 M_x R_x^2}{R^4 V^2}.$$

The time of the free path between collisions is $t_f = 1/(\pi R^2 nV)$, where $n$ is the number density of objects.

Introducing the effective collision frequency $v_E$ and the tidal destruction cross-section $\sigma_E$ according to the relations

$$\left\langle \frac{dE}{dt} \right\rangle = \frac{E}{t_f} = v_E E, \qquad v_E = nV\sigma_E \qquad (34)$$

($E$ is the absolute value of the total energy), we eventually find

$$\sigma_E = \pi R_x^2 \left(\frac{M}{M_x}\right)^2 \frac{R_x^2}{R^2} \frac{v_x^2}{V^2}, \qquad R \geqslant R_x, \qquad (35)$$

where $v_x$ is the mean velocity of dark-matter particles inside the objects,

$$\frac{M_x v_x^2}{2} = E = \frac{GM_x^2}{R_x}.$$

From Eqn (35) it can be seen that the maximum cross-section is achieved for $R \approx R_x$. Let us now take account of the fact that the velocity $v_x$ is determined by the gravitational potential of a considered object $M_x$, while the velocity $V$ is determined by the gravitational potential of a much larger-scale formation in which it is trapped. Consequently, as indicated above, we always have $V^2 \gg v_x^2$. Hence, the cross-section of tidal destruction of objects by one another is always rather small as compared to their geometric size. Thus, the lifetime of the objects is substantially larger than the time of their free path between straightforward head-on collisions.

As is clear from Eqns (34), the lifetime of objects of mass $M_x$ trapped in a large-scale object of mass $M_f$, i.e., the tidal destruction time $t_f$, is determined by the relation

$$t_f = \frac{1}{n(r)\sigma_E V_f}. \qquad (36)$$

Here, the concentration of trapped objects $n(r)$ dependent on the distance $r$ to the center of the large-scale object is given by relations (32) and (33) and the velocity $V_f$ is equal to

$$V_f = \sqrt{2G\frac{M_f}{R_f}}. \qquad (37)$$

In expression (35), for the cross-section one can put $M = M_x$, $R = R_x$. Proceeding from relations (35) – (37), we find

$$t_f = \frac{4R_f^2}{(3-\alpha)R_x f} \sqrt{\frac{R_f}{2GM_f}} \left(\frac{r}{R_f}\right)^\alpha. \qquad (38)$$

Here $f$ is the portion of the mass of the large-scale object $M_f$ contained in the small-scale entrapped objects $M_x$. For example, the total number of entrapped objects of mass $M_x$ is $N_x = fM_f/M_x$. It can be seen that the lifetime of entrapped objects decreases strongly with decreasing $r$, that is, approaching the center. On average, it increases proportionally to the object size ratio $R_f/R_x$ as well as with increasing characteristic time of oscillations in the large-scale object and with decreasing parameter $f$.

### 3.3 Destruction of objects in the hierarchical structure
The origination of the hierarchical structure is closely related to the Hubble expansion of the Universe. Indeed, because of the expansion, the mean density of matter $\rho_0 \propto t^{-2}$ falls rapidly. As a result, those objects that appeared earlier have a higher density. Therefore, they are more strongly gravita-tionally bound and can exist inside a large-scale object. It is the Hubble expansion and the sequence of appearance of objects — first small-scale, and then increasingly large-scale — that enabled the formation of the hierarchical structure. The sequence of appearance of objects, as was shown above, is specified by the form of the spectral function $F(k)$ (Fig. 2): small-scale objects appear earlier than large-scale ones if the function $F(k)$ grows monotonically with increasing $k$. For the standard Zel'dovich – Harrison spectrum this is not the case because the function $F(k)$ over a broad range of scales for $k > k_{eq}$ is nearly constant (Fig. 2, curve 1). In this case, objects of all sizes for $k > k_{eq}$ appear almost simultaneously. But given this, smaller objects inside larger ones are in no way distinguished in density. This means that they cannot originate through a regular process, and if they are distinguished owing to fluctuations, their density differs very little from the mean density of the surrounding matter, and they are destroyed rapidly.

From this it follows that in the case of a standard spectrum (curve 1 in Fig. 2) the hierarchical structure does not at all depend on the spectrum cut-off at small scales. It begins in the region where a deviation appears in the flat spectral function $F(k)$, that is, for $k \leqslant k_{eq}$. The scale $k_{eq}^{-1}$ is smaller than the horizon radius by approximately an order of magnitude at the moment of equilibrium onset. It corresponds in mass to small galaxies. Consequently, in the case of a standard spectrum the hierarchical structure begins in the region of small galaxies and stretches towards larger scales. This agrees well with the observed large-scale structure of the Universe.

The picture is approximately the same for the power-law spectrum which differs from the Zel'dovich – Harrison spectrum ($n = 1.4$, curve 2 in Fig. 2). Although the hierarchical structure may in this case start developing early, the survival small-scale objects to survive is hampered because of the small difference in density. A sufficiently long-lived structure occurs on a scale larger than $k_{eq}^{-1}$ when the slope of the spectral function $F(k)$ changes sharply. This again leads mainly to the possibility of the appearance of a full hierarchical structure in the Universe beginning with small galaxies.

The situation is essentially different in the case of the non-standard function $F(k)$ represented in Fig. 2 (curve 3). The hierarchical structure develops here first in the region of small scales, for $k > k_s$, but it breaks sharply for $k \leqslant k_s$ and appears again only on galactic scales $k \leqslant k_{eq}$. Small-scale objects $k \geqslant k_s$ originate early and have, therefore, a very high dark-matter density in galactic-scale objects that arise much later. Under such conditions small-scale dark-matter objects with $k \geqslant k_s$ are strongly pronounced. As will be seen below, they may exist in the Universe for a rather long time.

### 3.4 Small-scale structure — the main hypothesis
Sufficiently reliable data on the spectrum of initial perturbations in the Universe are now available only for large scales $l > 1 - 10$ Mpc. As to small scales $l < 1 - 10$ Mpc, the information is next to null. The basic hypothesis that we suggested in Ref. [13] reads that for small scales the spectral function exhibits a non-standard nature as shown in Fig. 2 (curve 3) and for this reason the small-scale structure develops in the Universe. On the assumption that it is this structure that is observed in the microlensing on objects in the halo of the Galaxy one may conclude that objects of mass $0.1 - 0.5 M_\odot$ correspond to the point of a sharp small-scale ($k_s$) spectrum cut-off. Furthermore, the amplitude of the spectral function in the plateau (or peak) region is

$$F_m = \delta_{im} \approx 0.3 - 0.4, \qquad (39)$$

which exceeds the value of the constant $F_0$ in the standard spectrum by over two orders of magnitude.

On this basis we obtain that the small-scale hierarchical structure of dark-matter objects extends to the region

$$M_x \sim (0.01 - 1) M_\odot. \qquad (40)$$

Since the initial amplitude is high (39), it follows that as is implied by Eqn (20), the fluctuations reach a nonlinear stage and form gravitationally compressed objects within a period of time close to the moment of equilibrium onset $t_{eq}$. The scale of the appearing objects (22) and (23) is

$$R_x \sim (10^{14} - 10^{15}) \text{ cm}. \qquad (41)$$

This scale grows with increasing mass $M_x$ in proportion to $M_x^{1/3}$ or somewhat faster, the growth rate depending on the slope of the spectral function.

We shall emphasize that the principal hypothesis of the character of the spectrum of initial perturbations has not yet been cosmologically grounded.

### 3.5 Lifetime of small-scale objects in the Galaxy

The small-scale structure with the characteristic mass spectrum (40) and scales (41) develops, under assumption (39), over the period close to the moment of equilibrium onset $t_{eq}$. At this moment, the amplitude of fluctuations on large scales is only $F_0 \sim 10^{-3}$. The large-scale structure, therefore, develops much later, when $z < z_0$ ($z_0 \leqslant 10$). As a result, the mean dark-matter density in large-scale structures, including the halo of our Galaxy, is many ($\sim 10^{10}$) orders of magnitude lower than the density in small-scale objects. The small-scale objects are thus strongly distinguished dense formations in the Galactic halo.

They, naturally, interact among themselves as well as with stars and galactic gas and can, in principle, decay as a result of this interaction. We shall determine their lifetime. To this end, we shall first estimate the total number of objects in the Galactic halo

$$N_x M_x = f M_h. \qquad (42)$$

Here $f$ is the fraction of dark matter contained in small-scale structures, $M_h$ is the total mass of the dark matter in the halo, $M_x$ is the characteristic mass of an object. The distribution of dark-matter objects obeys the fundamental law (1). Consequently, the density of these objects in the halo is

$$n_x(r) = \frac{3-\alpha}{4\pi} N_x R_h^{-3} \left(\frac{r}{R_h}\right)^{-\alpha}, \qquad (43)$$

where $R_h$ is the size of the halo.

Allowing for the tidal interaction of the dark-matter objects, their lifetime is determined by expression (38). Substituting Eqns (42) and (43), we obtain

$$\tau_0 = \frac{4 R_h^2}{(3-\alpha) R_x f} \sqrt{\frac{R_h}{2GM_h}} \left(\frac{r}{R_h}\right)^\alpha. \qquad (44)$$

Using now the values of the parameters [6, 13]

$$M_h \approx 2 \times 10^{12} M_\odot, \qquad R_h \approx 200 \text{ kpc},$$
$$M_x \approx 0.5 M_\odot, \qquad R_x \approx 4 \times 10^{14} \text{ cm}, \qquad f \approx 0.5,$$

we find

$$\tau_0 = 7 \times 10^8 t_0 \left(\frac{R_h}{200 \text{ kpc}}\right)^{3/2} \left(\frac{4 \times 10^{14} \text{ cm}}{R_x}\right)$$
$$\times \sqrt{\frac{2 \times 10^{12} M_\odot}{M_h}} \left(\frac{r}{200 \text{ kpc}}\right)^{1.8}. \qquad (45)$$

Here $t_0 \approx 3 \times 10^{17}$ s is the lifetime of the Universe. From this one can see that the lifetime of objects in the halo is many orders of magnitude larger than the lifetime of the Universe. Approaching the central part of the halo, $\tau_0$ decreases substantially, but here, too, below scales of the order of 10 kpc it remains much greater than $t_0$.

In the galactic region, one should also involve the interaction with stars. Making use of relations (35), (36), we arrive at the lifetime

$$\tau_s \approx 4 \times 10^3 t_0 \left(\frac{4 \times 10^{14} \text{ cm}}{R_x}\right) \left(\frac{V}{300 \text{ km s}^{-1}}\right)$$
$$\times \left[\frac{1 \text{ pc}^{-3}}{N_s(r)}\right] \left(\frac{M_x}{0.5 M_\odot}\right) \left(\frac{M_\odot}{M_s}\right)^2, \qquad (46)$$

where $N_s(r)$ is the density of the stars and $M_s$ is their characteristic mass. We can see that outside the central region $r > r_{Gc}$, where $r_{Gc} \sim 0.1 - 1$ kpc, this quantity remains larger than the lifetime $t_0$.

Thus, dark-matter objects may exist at the present time not only in the halo, but in the greater part of the Galaxy as well. It is in the region $r < r_{Gc}$ close to the center that they must already be destroyed as a result of tidal interaction with stars.

The interaction of dark-matter objects with gas may be quite diversified. First of all, since they form before recombination, it follows that after recombination and gas cooling to temperatures

$$T < T_0, \qquad T_0 \approx \frac{m_p M_x G}{R_x} \approx 2.6 \times 10^3 K, \qquad (47)$$

where $m_p$ is the proton mass, gas condensation starts in the dark-matter objects, which leads to the formation of baryonic cores. This process may proceed even now. Then, obviously, a substantial difference must appear in scale and, perhaps, in the composition and structure of baryonic cores in the halo and the Galaxy, where the gas density is high. Baryonic bodies in the Galaxy may not only be much greater than in the halo, but they may also have an appreciable effect on the distribution of the dark matter surrounding the core. Specific features of the structure of dark-matter objects with baryonic cores will be discussed at length in the section to follow.

Note that being bare gravitational bodies on which star-constituting gas is condensed, the dark-matter objects existing in the Galaxy may also have a strong effect on the star formation process. On birth and ignition of a star, dark-matter particles may either leave it or stay inside (depending on their mass $m_x$) producing a noticeable effect on its burning. This point will also be discussed briefly in Section 4.

Concluding this section we emphasize that the small-scale structures considered within the proposed theory are essentially new objects in the structure of matter in the Universe. They are formed by non-interacting dark-matter particles and exist owing to gravitational forces only, i.e., the particles are entrapped by the gravitational field created by these objects.

An important feature of the objects is, as follows from Eqns (40) and (41), that their mean density is $\rho \propto 10^{-11} - 10^{-13}$ g cm$^{-3}$. In view of this, in spite of a similar mass, the volume they occupy is much larger than is typical for baryonic bodies — their size exceeds that of a corresponding compact baryonic body by approximately four orders of magnitude. For this reason we henceforth refer to them as non-compact objects (NO). It is noteworthy that, having a similar mass, a baryonic gas occupies a much larger (by 7–9 orders of magnitude) volume in the Galaxy than NO. Thus, one can state that in their general properties — mass, scale, and luminosity — NO are specific objects of non-baryonic matter. Weakly luminous baryonic objects of the same mass and size do not exist in nature.

## 4. Structure of non-compact objects

### 4.1 Coefficient of nonlinear compression

In the course of gravitational collapse leading to the formation of gravitationally bound objects of non-interacting dark-matter particles, nonlinear compression occurs. Let us determine the parameters of this compression. Gravitationally compressed objects are formed in the neighborhood of local maxima of the initial density. Suppose at the initial moment the density distribution near this maximum $r = 0$ has the form

$$\rho = \rho_0 \left(1 - \frac{r^2}{r_0^2}\right), \quad r < r_0. \tag{48}$$

For simplicity, we have assumed here the initial distribution to be symmetric (in the neighborhood of a three-dimensional maximum of any form the results are quite similar [5, 31]). The evolution of the nonlinear process of gravitational compression of the initial cluster (48) leads to the appearance in a time of the order of the Jeans time ($t_J \simeq \pi\sqrt{3/8\rho}$) of a singular point in the center with a density distribution [4]

$$\rho(r) = Kr^{-12/7}, \quad K = \frac{3}{7}\left(\frac{40}{9\pi}\right)^{6/7} \rho_0 r_0^{12/7}. \tag{49}$$

Let us assume the distribution (49) to be restricted to the radius $r_1$. For $r > r_1$ we put $\rho = 0$. The forms of the distributions (48) and (49) differ strongly. Under these conditions, in order to find the value of the effective parameter of compression it is reasonable to bring both the distributions to a unified form. It is natural to choose as such a form a uniform ball with a constant density

$$\rho = \rho_{\text{uni}}. \tag{50}$$

To the initial distribution (48), we add a uniform ball of the same mass

$$M = \frac{8}{15}\pi\rho_0 r_0^3 = \frac{4}{3}\pi\rho_{\text{uni}}^{(0)} r_0^3 \tag{51}$$

and radius $r_0$. Then the density $\rho_{\text{uni}}^{(0)}$ is related to the parameter $\rho_0$ as follows

$$\rho_{\text{uni}}^{(0)} = \frac{2}{5}\rho_0. \tag{52}$$

A uniform ball of radius $r_1$ and density $\rho_{\text{uni}}^{(1)}$ corresponds to the singular density distribution (49). From Eqns (49) we obtain

$$\frac{4}{3}\pi\rho_{\text{uni}}^{(1)} = \frac{28\pi}{9} K r_1^{-12/7}. \tag{53}$$

Furthermore, in view of mass conservation we have the relation

$$\rho_{\text{uni}}^{(0)} r_0^3 = \rho_{\text{uni}}^{(1)} r_1^3. \tag{54}$$

From Eqns (49), (52), and (53) we find the coefficients of compression in density $\varkappa_\rho$ and in radius $\varkappa_r$:

$$\varkappa_\rho = \frac{\rho_{\text{uni}}^{(1)}}{\rho_{\text{uni}}^{(0)}} \simeq \left(\frac{40}{9\pi}\right)^2 \left(\frac{5}{2}\right)^{7/3}, \quad \varkappa_r = \frac{r_1}{r_0} = \varkappa_\rho^{-1/3}, \tag{55}$$

which gives $\varkappa_\rho \simeq 20$ and $\varkappa_r \simeq 0.3$. Let us emphasize that the coefficients of compression (55) do not depend on a particular unified form to which both the distributions are brought. For example, the result will be the same if we bring the distribution (49) to the form (48) assuming $\rho(r) = \rho_1(1 - r^2/r_1^2)$ and determine the parameters $\rho_1$ and $r_1$ using mass conservation.

We can see that the nonlinear gravitational compression is sufficiently large even before the first singularity. The analysis carried out in Ref. [32] shows that in the course of subsequent mixing and the onset of the stationary state it does not generally increase. This remains valid if all the particles are kept in the entrapment region. However, depending on the initial conditions, some particles may actually leave the entrapment region, and as a result of conservation of the total energy this process inevitably leads to strengthening the object's compression.

### 4.2 Density distribution in non-compact objects

The distribution of density in NO depends mainly on the development of Jeans instability at the nonlinear stage, but the linear stage of fluctuation growth also exerts a considerable influence on this process. So, the scaling law (1), (49) is characterized by two factors, namely, the slow linear growth of initial density fluctuations and the rapid compression with subsequent kinetic mixing at the nonlinear stage. In the case of large scales, the initial density fluctuations $\delta_{i0}$ [see Eqns (2) and (3)] are always small, and therefore the unstable mode increases for a substantially long time before the nonlinear collapse sets in. Within this time interval, the stable modes are damped strongly. In the course of nonlinear collapse, however, one of the damping modes starts increasing and does so very rapidly, so that in the neighborhood of $r = 0$ it outruns the increasing mode [33]. As a result, in a small neighborhood of $r = 0$ the law (1), (49) is cut off on the scale $r_c$ [5], where

$$r_c = \delta_{i0}^3 R_x. \tag{56}$$

Here $R_x$ is the characteristic scale of an object. The magnitude of initial density fluctuations $\delta_{i0}$ on large scales is rather small, $\delta_{i0} \simeq 10^{-3}$. This means that the scaling law (49) is fulfilled up to very small scales: from Eqn (56) it follows that $r_c \sim 10^{-9} R_x$. Thus, its actual cut-off is naturally caused by the influence of other processes, for instance, by the action of baryonic matter or the appearance of a giant black hole [34].

For small-scale objects the situation is essentially different. As shown above [see Eqn (39)], the initial density fluctuations on small scales, where their effective growth begins, are sufficiently high: $\delta_{i0} \sim 0.3-0.5$ for $t = t_{eq}$. Then from Eqn (56) it follows that the distribution (1), (49) is cut off rather early, namely when

$$r_c \sim (0.05-0.1)R_x. \quad (57)$$

As a result, the particle density in the CDM of a small-scale NO can approximately be represented as

$$\rho = \begin{cases} \rho_0, & 0 < r < r_c, \\ \rho_0 \left(\dfrac{r}{r_c}\right)^{-\alpha}, & r_c < r < R_x, \\ 0, & r > R_x, \end{cases} \quad (58)$$

where the scale $r_c$ is determined according to Eqn (57) and the density $\rho_0$ is related to the object's mass $M_x$ and the scale $R_x$ as

$$\rho_0 \simeq \frac{3-\alpha}{4\pi} M_x R_x^{\alpha-3} r_c^{-\alpha}. \quad (59)$$

### 4.3 Influence of baryonic component on the structure of non-compact objects

Up to now we have ignored the influence of the small baryonic component on the structure of dark-matter in NO. However, as they cool down, baryons tend to the bottom of the potential well created by the dark matter and condense at the center forming a compact baryonic object of mass $M_b$. It creates an additional potential

$$\psi_b = -\frac{GM_b}{r}.$$

The scale of influence of this potential, $r_b$, depends on the mass of a baryonic object and can readily be estimated on the basis of the relation

$$\psi_b(r_b) = \psi_d(r_b), \quad (60)$$

where $\psi_d(r)$ is the potential created by the non-baryonic component with the density distribution $\rho_d(r)$ (58), (59). It can be found through a solution of the Poisson equation

$$\frac{1}{r^2}\frac{d}{dr}\left(r^2 \frac{d\psi_d}{dr}\right) = 4\pi G \rho_d(r) \quad (61)$$

with the boundary condition $\psi_d \to 0$ as $r \to \infty$; it has the form

$$\psi_d = \begin{cases} \dfrac{GM_x}{R_x}\left\{\left[\dfrac{3-\alpha}{6}\left(\dfrac{r}{r_c}\right)^2 + \dfrac{\alpha(3-\alpha)}{2(2-\alpha)}\right]\right. \\ \quad \times \left(\dfrac{r_c}{R_x}\right)^{2-\alpha} - \dfrac{3-\alpha}{2-\alpha}\Bigg\} \\ \quad \times \left[1 - \dfrac{\alpha}{3}\left(\dfrac{r_c}{R_x}\right)^{3-\alpha}\right]^{-1}, & r < r_c; \\ -\dfrac{GM_x}{r} + \dfrac{GM_x}{R_x}\left[\dfrac{1}{2-\alpha}\left(\dfrac{r}{R_x}\right)^{2-\alpha}\right. \\ \quad \left. -\dfrac{3-\alpha}{2-\alpha}+\dfrac{R_x}{r}\right]\left[1-\dfrac{\alpha}{3}\left(\dfrac{r_c}{R_x}\right)^{3-\alpha}\right]^{-1}, & r_c \leqslant r \leqslant R_x; \\ -\dfrac{GM_x}{r}, & R_x \leqslant r; \end{cases}$$

$\alpha = 1.8. \quad (62)$

From relations (60) and (62) it follows that in the case $M_b \sim 0.05 M_x$ the region of appreciable influence of the baryonic-object potential is of the order of $r_b \approx (0.05-0.1)R_x$. For smaller values of $M_b$ the region of influence of the potential $\psi_b$ is even smaller.

We shall assume the time of oscillations of dark-matter particles trapped in the potential (62) to be much smaller than the characteristic time of baryon cooling and baryonic body formation in the center of NO. For the objects of interest, (40) and (41), the time of oscillations is $t_0 \sim 10$ years, so this condition is always well fulfilled. In this case, to consider the variation of the distribution function of dark-matter particles under the influence of baryons, one can use the adiabatic approximation. The initial adiabatic invariant $I_i$ is determined by the relation

$$I_i = \int_{r_{\min}}^{r_{\max}} \sqrt{E - \psi_d(r) - \frac{m^2}{2r^2}}\, dr, \quad I_i = I_i(E,m). \quad (63)$$

Here $E$ is the energy of a particle moving in the potential $\psi_d$, normalized to the mass $m_x$; $m$ is the angular momentum of the particle; $r_{\min}$, $r_{\max}$ are reflection points defined as points at which the integrand vanishes. The initial distribution function is

$$f = f_0(I_i) = f_0(E,m). \quad (64)$$

After the baryonic body formation, the adiabatic invariant $I_b$ of the particle is already described by another equation

$$I_b = \int_{r_{\min}}^{r_{\max}} \sqrt{E - \psi(r) + \frac{GM_b}{r} - \frac{m^2}{2r^2}}\, dr, \quad (65)$$

where $GM_b/r$ is the baryonic body potential and $\psi(r)$ is the potential created by the non-baryonic matter. In view of the fact that the adiabatic invariant is preserved in a slow process, $I = \text{const} = I_b$, the relation

$$f(I) = f_0(I_i)\Big|_{I_i = I_b} \quad (66)$$

holds. Here $f_0(I_i)$ is the initial distribution function, but given this, the function $I_b(E,m)$ (65) already differs from $I_i(E,m)$ (63). Expressions (63)–(66) completely describe the deformation of the distribution function $f(E,m)$ of dark-matter particles under the influence of the baryonic body potential. The distributions of the potential and particle density are described here by the equations

$$\Delta \psi = 4\pi G \rho,$$
$$\rho(r) = \frac{2\pi m_x}{r^2}\int_0^\infty dm^2 \int_{\psi+m^2/(2r^2)}^0 \frac{f(E,m)}{\sqrt{E-\psi-m^2/(2r^2)}}\, dE. \quad (67)$$

To determine the NO structure, one should solve Eqns (67) together with Eqns (63)–(66). We shall stress that these equations, which determine the distribution function $f(E,m)$, the potential $\psi(r)$, and the dark-matter density distribution $\rho(r)$, are essentially nonlinear. The perturbations of the potential $\psi$ and density $\rho$ by the baryonic body depend, naturally, on the mass of this body. We shall first analyse (Section 4.4) the case of a small-mass baryonic body $M_b/M_x \to 0$, when the perturbation of the potential $\psi$ created by non-baryonic matter can be neglected to a first approximation under the assumption that $\psi \simeq \psi_d(r)$. After that we shall consider (Section 4.5) the influence of a baryonic body of arbitrary mass.

### 4.4 Structure of non-compact objects in the halo

As was shown above, when the baryonic body mass is not large, $M_b \ll M$, the region of its influence is $r_b < r_c$ (57). Hence, in solving Eqns (67) one can restrict the consideration to the region $r \leqslant r_c$ (58). In this region, the distribution function should naturally be chosen in the Maxwellian form

$$f_0 = \frac{n_0}{(2\pi T)^{3/2}} \exp\left(-\frac{E}{T}\right), \qquad T \approx \phi = \frac{2}{3}\pi G n_0 m_x r_c^2. \quad (68)$$

Here $\phi$ is the potential well depth at the level $r = r_c$. On integrating equations (63) and (65), we obtain the following relations for adiabatic invariants

$$I_i = \frac{\pi}{2^{3/2}}\left(\sqrt{\frac{2}{\phi_0}}E - m\right), \qquad I_b = \frac{\pi}{2}\left(\frac{2GM_b}{(-E)} - 2^{1/2}m\right). \quad (69)$$

After the substitution of Eqns (68) into (67), the expression for the density has the form

$$\rho(r) = \frac{2\pi m_x}{r^2}\int_0^\infty dm^2 \int_{-GM_b/r + m^2/(2r^2)}^0 \frac{n_0}{(2\pi T)^{3/2}}$$
$$\times \exp\left(\frac{2\phi_0^{1/2}GM_b}{TE} + \frac{m\phi_0^{1/2}}{2^{1/2}T}\right)\left(E - \psi - \frac{m^2}{2r^2}\right)^{-1/2} dE. \quad (70)$$

Let us analyse relation (70) in the neighborhood of a baryonic body, i.e., as $r \to 0$. In this case we have $E \to -\infty$ and allowing for the fact that $m^2 \leqslant 2GM_b r$ we find

$$\rho(r) = \frac{8}{3\sqrt{2\pi}}\hat{n}_0 m_x \left(\frac{GM_b}{Tr}\right)^{3/2}. \quad (71)$$

Note that although the result (71) is obtained under the assumption (68) concerning the form of the initial distribution function, the asymptote (71) does not depend on the form of the function but is determined only by its behavior as $E \to -\infty$. It is the value of the constant $\hat{n}_0$ that depends on a particular form of the distribution function. As is seen from Eqn (71), the density distribution has a singularity when $r \to 0$, but this singularity is due to the fact that in the calculations we assumed the baryonic body to be a point. Therefore, the law (71) will actually be cut off on the baryonic body scale $r_b$. Proceeding from these considerations and allowing for the fact that, as follows from relations (57) and (58),

$$\frac{4}{3}\pi n_0 m_x r_c^3 \approx 0.1 M_x,$$

we find the density of dark-matter particles trapped in the baryonic body:

$$\rho(r_b) \approx 32 n_0 m_x \left(\frac{M_b r_c}{M_x r_b}\right)^{3/2}. \quad (72)$$

From Eqns (72), (57), (40), and (41) it follows that an adiabatic entrapment causes an increase of dark-matter particle density by three to four orders of magnitude both inside the baryonic body and in its neighborhood (71). For a NO mass of the order of $0.5 M_\odot$, the total mass of dark-matter particles trapped in the baryonic core appears to be

$$M_n \sim 10^{25} - 10^{26} \text{ g}. \quad (73)$$

### 4.5 Non-compact objects in the Galactic disc

As has already been mentioned above, NO in the galactic disc are centers of interstellar gas condensation, and owing to this fact their baryonic core can increase substantially. Such an increase of the baryonic core affects the general dark-matter distribution and can even change the scale of the NO.

When the fraction of baryonic matter becomes significant, in the solution of Eqns (67) and calculation of the adiabatic invariant (65), it is necessary to take into account the potential of the dark and baryonic matter over the entire solution space. Here we shall restrict ourselves to examination of the coefficient of dark-matter compression $\varkappa_{dm}$ which we define according to the relation

$$\varkappa_{dm} = \frac{\langle\rho_i\rangle}{\langle\rho_f\rangle}, \quad (74)$$

where $\langle\rho_i\rangle$ is the mean dark-matter density in the NO before and $\langle\rho_f\rangle$ is the same after the baryonic body formation. In view of mass conservation, by analogy with Eqn (54) we obtain the relation of $\varkappa_{dm}$ with the body size ratio:

$$\varkappa_{dm} = \frac{R_i^3}{R_f^3}. \quad (75)$$

To find the body size ratio, we shall make use of conservation of the adiabatic invariant. We shall determine the maximal adiabatic invariant by setting $E = -GM_x/R_x$ in Eqn (63):

$$I^{max} = I\left(E = -\frac{GM_x}{R_x}\right).$$

Taking into account that according to the general theory [4] the moment $m$ is small, we find from Eqn (63) that in the initial state

$$I_i^{max} = 0.37\sqrt{5GM_x R_i}, \qquad R_i = R_x. \quad (76)$$

In the presence of a baryonic body we find the final value of the adiabatic invariant from Eqn (65)

$$I_f^{max} = I\left(E = -\frac{GM_x}{R_f}\right) = \frac{\pi}{2}\frac{M_b}{M_x}\sqrt{\frac{GM_x R_i}{R_i/R_f - 1}}. \quad (77)$$

Equating these two invariants, we find the degree of NO compression on the scales:

$$\frac{R_i}{R_f} = 1 + 3.6\left(\frac{M_b}{M_x}\right)^2. \quad (78)$$

From Eqns (75) and (78) one can see that the degree of NO compression depends strongly on the $M_b$-to-$M_x$ ratio. So, for $M_b = 0.3 M_x$ the mean dark-matter density increases by a factor of 2.5 and for $M_b = 0.5 M_x$ by a factor of 7. As is seen in the sequel (see Section 6), such a compression of a NO may have a substantial effect upon its gamma-radiation.

## 5. The theory of microlensing on non-compact objects

The possibility of rays of light being focused by a gravitational field (lensing) was first reported by O Chwolson [35] and was later considered by many authors [36, 37]. The gravitational lensing phenomenon was first observed for the radiation of quasars, and the lensing objects in that case were galactic halos [38–40].

An important step was made by B Paczynski who proposed the observation of star light lensing by dark-matter objects in the halo of our Galaxy (supposedly Jupiter type planets) [17]. This work underlay the experimental studies [14, 41] which led to the discovery of the microlensing effect. We shall describe here the theory of microlensing on compact and non-compact bodies.

### 5.1 Microlensing on compact bodies

Let us consider gravitational lensing in the case where the lensing object is compact, i.e., its size is negligibly small. The potential created by an object of mass $M$ is then written in the form

$$\Phi_s(r) = -\frac{GM}{r}. \tag{79}$$

The equation of the trajectory of a light ray in a spherically symmetric gravitational field

$$ds^2 = \exp[\nu(r)]c^2 dt^2 - \exp[\lambda(r)] dr^2 - r^2(d\theta^2 + \sin^2\theta\, d\phi^2),$$

where $\nu(r)$ and $\lambda(r)$ are arbitrary functions, has the form (see Refs [42, 43])

$$\left(\frac{du}{d\phi}\right)^2 + u^2 \exp(-\lambda) - \frac{1}{R^2}\exp(-\nu - \lambda) = 0, \qquad u = r^{-1}. \tag{80}$$

Here $R$ is the incidence parameter of the ray and the angle $\phi$ is taken in the ray plane (Fig. 3).

In the zeroth approximation in $\nu, \lambda$ the solution to Eqn (80) is the straight line

$$u^{(0)} = \frac{1}{R}\cos(\phi - \phi_0). \tag{81}$$

The first correction gives

$$u^{(1)} = \frac{1}{2R}\left\{\int^{\phi}\left[\frac{\nu + \lambda}{\sin^2(\phi' - \phi_0)} - \frac{\lambda\cos^2(\phi' - \phi_0)}{\sin^2(\phi' - \phi_0)}\right]d\phi' + C_1\right\}\sin(\phi - \phi_0), \tag{82}$$

the integral being calculated along the straight line (81). In Newton's approximation we have

$$\nu(r) = -\lambda(r) = \frac{2\Phi(r)}{c^2}. \tag{83}$$

The calculation of the angle between the asymptotes yields the total deviation of the ray

$$\Theta = \int_0^{\pi/2}\left[\lambda(r') - r'\lambda'_r(r')\right]d\phi', \qquad r'(\phi') = \frac{R}{\cos\phi'}. \tag{84}$$

Substituting the potential (79) into Eqns (84) and (83) and integrating, we arrive at the Schwarzschild value $\Theta = \Theta_s(R)$:

$$\Theta_s(R) = \frac{4GM}{c^2 R}. \tag{85}$$

Thus we have found the dependence $\Theta(R)$ of the angle of ray deviation from the incidence parameter in the given potential (79).

Let us now turn to the mutual position of an observer, a lens, and a source of light (Fig. 4). Suppose $\theta$ is the angle between directions from the observer to the lens D and to the source of light (a star) S, $L_D$ and $L_S$ are the distances to the lens and to the star, respectively, $L_{SD} = L_S - L_D$. Because the rays of light are bent, the observer will see two images $I_1$ and $I_2$ instead of the point star in the ODS plane. By $\theta_1$ we shall denote the angle between the directions to the lens and to the image. Then, from geometric considerations (see Fig. 4 and, e.g., Ref. [44]) and allowing for the smallness of the angles $\theta$, $\theta_1$, $\Theta$, one can obtain the formula

$$\theta_1 \pm \theta = \frac{L_{SD}}{L_S}\Theta(R), \qquad R = \theta_1 L_D \tag{86}$$

which relates the angles $\theta$ and $\theta_1$. The sign $\pm$ is taken to make all the angles positive.

When the Schwarzschild function (85) is substituted into Eqn (86), a dimensionless quantity appears, namely,

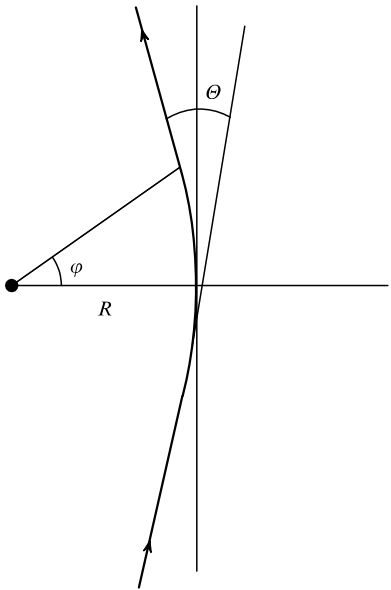

**Figure 3.** Trajectory of a light beam in a spherically symmetric gravitational field.

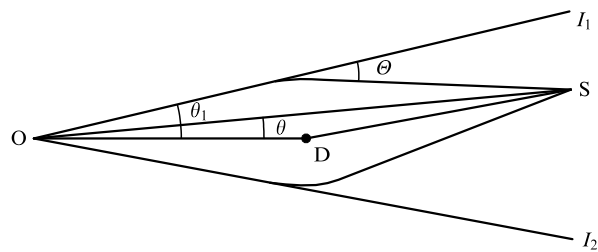

**Figure 4.** Geometry of a gravitational lens. (Mutual positions of an observer O, a lens D and a source S.)

the angle $\theta_0$

$$\theta_0^2 = \frac{4GM}{c^2} \frac{L_{\text{SD}}}{L_S L_D} \quad (87)$$

or the Einstein radius

$$R_E^2 = (L_D \theta_0)^2 = \frac{4GM}{c^2} \frac{L_{\text{SD}} L_D}{L_S}. \quad (88)$$

The relation (86) itself in this case passes over to the quadratic equation

$$\theta_1(\theta_1 \pm \theta) = \theta_0^2, \quad (89)$$

which always has two solutions $\theta_1(\theta)$.

The Einstein radius gives the characteristic lens scale. The lensing object is compact if its size $r_b$ is much smaller than $R_E$. For planets or stars on scales of the Galactic halo this condition is always well fulfilled.

The dependence of $\theta_1^{\pm}$ on the angle $\theta$ is presented in Fig. 5. It reflects the variation of angular position of images $\theta_1^{\pm}$ as a function of the angle $\theta$ between the directions from the observer to the lens and to the source. As $\theta \to 0$, the images merge. As $\theta \to \infty$, one of the images approaches the star and the other approaches the lens. (As is seen from what follows, the brightness of the second image will tend to zero.)

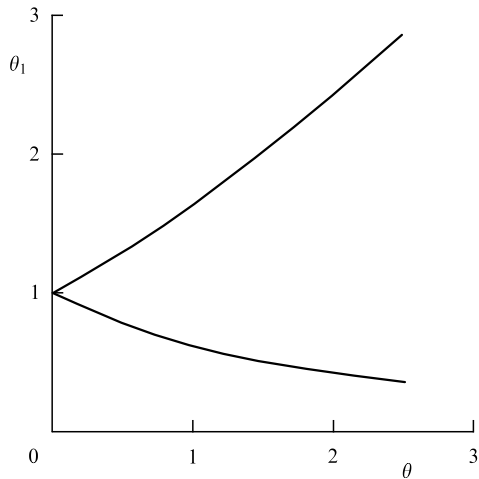

**Figure 5.** Position of images $\theta_1(\theta)$ in the case of a point-like lens. Angles are normalized to the Einstein angle $\theta_0$ (87).

It is of importance that under the conditions of experiments on microlensing the characteristic distance to a lensing object in the Galactic halo is $L_D \sim 10$ kpc. If the mass of this object is $M \sim 1 M_\odot$, the angle between the images is $\theta_1 \sim 0.001''$. Such angles are too small to allow image resolution using present technical means. But the existence of gravitational focusing can be established from brightness variation of the lensed star.

The brightness amplification coefficient $\Omega$ is equal to the ratio of the total angular area of the images to the area of the source. It is expressed in terms of the position of source and images as follows

$$\Omega = \sum \frac{\theta_1}{\theta} \left| \left(\frac{d\theta_1}{d\theta}\right)_\theta \right|. \quad (90)$$

Here the sum is taken over both branches $\theta_1(\theta)$.

With allowance for Eqn (89), we obtain

$$\Omega_s = \frac{\theta^2 + 2\theta_0^2}{\theta \sqrt{\theta^2 + 4\theta_0^2}}. \quad (91)$$

Let us now take into account that the picture of star brightness amplification is non-stationary: the observer, the lens D and the star S possess some virial velocities. In view of this, the angle $\theta$ varies with time. Since it is the relative velocity alone that is significant, one may assume that the observer and the star are at rest, while the lensing object is moving at a velocity $v_\perp$ in the plane perpendicular to the line of sight. Given this, the angle $\theta$ between the directions to the source S and lens D is equal to

$$\theta = \sqrt{\theta_{\min}^2 + \left[\frac{v_\perp(t - t_{\min})}{L_D}\right]^2},$$

where $t_{\min}$ is the instant when they are maximally close to each other and $\theta_{\min}$ is the angular distance between D and S (incidence parameter) attained at this instant.

Finally, making use of Eqn (91) we obtain the time dependence of the star brightness amplification coefficient [15]:

$$A_s(t) = A_s[u(t)] = \frac{u^2 + 2}{u\sqrt{u^2 + 4}},$$
$$u(t) = \left[u_{\min}^2 + \left(\frac{2(t - t_{\min})}{\hat{t}}\right)^2\right]^{1/2}. \quad (92)$$

Here $u(t) = \theta(t)/\theta_0$, $u_{\min} = \theta_{\min}/\theta_0$, $\hat{t}$ is the characteristic lensing time related to $v_\perp$ as follows

$$\hat{t} = \frac{2R_E}{v_\perp} = \frac{2L_D \theta_0}{v_\perp}.$$

The relative motion of the lens and the source corresponds to the motion of the points in the graph $\Omega_s(x)$ from $x \to +\infty$ to $x_{\min}$ and back. Therefore, $A_s(t)$ is a symmetric 'bell-like' function which has a maximum value

$$A_{s,\max} = \Omega_s(\theta_{\min}),$$

and does not depend on the frequency of the light. It is the dependence $A(t)$ that is measured in observations.

Thus, in the case of a compact lens the shape of the curve $A(t)$ is completely determined by two parameters, namely, the incidence parameter $u_{\min}$ (or the $A$ value in the maximum) and the duration of the event $\hat{t}$. Figure 6 presents graphs of the dependence $A(t)$ for different values of the parameters $u_{\min}$ and $\hat{t}$; the instant of closest proximity $t_{\min}$ is assumed to be zero.

### 5.2 Microlensing on non-compact objects

The theory presented above refers exclusively to compact bodies whose scale $r_b$ is much smaller than the Einstein radius

$$r_b \ll R_E.$$

For non-compact dark-matter objects, the inverse condition typically holds

$$R_x \gtrsim R_E. \quad (93)$$

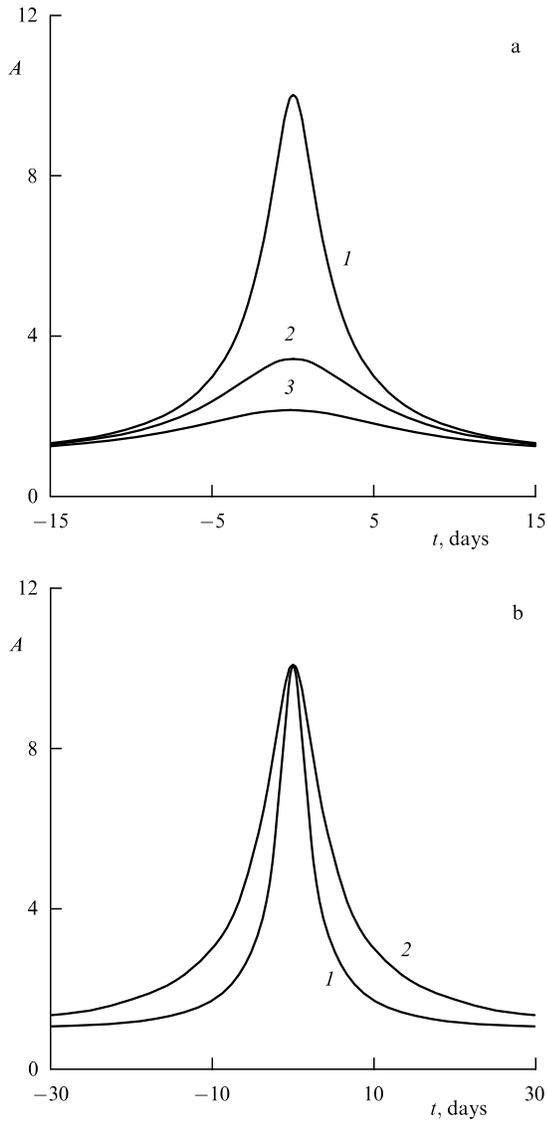

**Figure 6.** Dependence of the amplitude $A(t)$ of increase of for various $u_{\min}$, $\hat{t}$ in the case of a point-like lens: (a) $\hat{t} = 30$ days, $u_{\min}$ is equal to (*1*) 0.1, (*2*) 0.3, (*3*) 0.5; (b) $u_{\min} = 0.1$, $\hat{t}$ is equal to (*1*) 30, (*2*) 60 days.

Indeed, according to Eqn (23) the size $R_x$ of an object of mass $M \sim (0.1-1)M_\odot$ is of the order of $10^{14}-10^{15}$ cm. This exceeds by 3–10 times the Einstein radius for bodies of the same mass which are at a characteristic distance $L_D \sim 10$ kpc typical of lensing in the halo. Hence, the theory of microlensing should be extended to the case of NO with the characteristic sizes (93). It should be emphasized that such an extension of the microlensing theory is of interest for non-baryonic objects only because baryonic objects of such a scale and mass cannot exist in the Galaxy (see Section 3.5)

So, we shall consider gravitational lensing in the case where the lens size $R_x$ cannot be neglected. We shall assume the density distribution for $r \leqslant R_x$ to have the simplest form (1), i.e., the density cut-off for small values $r \leqslant r_c$ (58) is insignificant as shown below in Section 5.4. To the density distribution

$$\rho = \begin{cases} \dfrac{(3-\alpha)M_x}{R_x^3}\left(\dfrac{r}{R_x}\right)^{-\alpha}, & \alpha = 1.8, \quad r \leqslant R_x ; \\ 0, & r > R_x \end{cases} \tag{94}$$

corresponds the potential

$$\Phi(r) = \begin{cases} -\dfrac{GM_x}{R_x}\left[\dfrac{3-\alpha}{2-\alpha} - \dfrac{1}{2-\alpha}\left(\dfrac{r}{R_x}\right)^{2-\alpha}\right], & r \leqslant R_x ; \\ -\dfrac{GM_x}{r}, & r > R_x . \end{cases} \tag{95}$$

Bending of the trajectory of the ray of light for the spherically symmetric potential $\Phi(r)$ is described as before by formulae (80)–(84). For a complete deviation of the ray between the asymptotes, instead of Eqn (85) we now obtain

$$\Theta(R) = \dfrac{2GM_x}{c^2 R_x}\, f\!\left(\dfrac{R}{R_x}\right), \tag{96}$$

$$f(\xi) = \begin{cases} \dfrac{3-\alpha}{2-\alpha}\,\eta + 2\,\dfrac{1-\sin\eta}{\cos\eta} + \dfrac{1}{2-\alpha}\sin\eta\cos\eta \\ \quad -\dfrac{\alpha}{2-\alpha}\cos^{2-\alpha}\eta \displaystyle\int_0^\eta \cos^\alpha\phi'\, d\phi', & \xi \leqslant 1 ; \\ \dfrac{2}{\xi}, & \xi > 1 ; \end{cases}$$

$$\eta = \arccos\xi. \tag{97}$$

For $R > R_x$ the expression for $\Theta$ coincides with Schwarzschild's one.

The angle $\theta_1$ between the direction to the lens D and to the image is related as before to the position $\theta$ of the source S relative to the lens by formula (86),

$$\theta_1 \pm \theta = \dfrac{L_{SD}}{L_S}\,\Theta(R_1), \qquad R_1 = \theta_1 L_D .$$

However, the quantity $\Theta(R)$ is now determined not by Eqn (85) but by expression (96). Because the function $\Theta(R)$ has a complicated form, it is more convenient to transform the variables $\theta$, $\theta_1$ to the variables $x = \theta L_D/R_x$, $x_1 = \theta_1 L_D/R_x$ and to introduce instead of the Einstein angle $\theta_0$ the constant

$$Q = \dfrac{2GM_x}{c^2}\,\dfrac{1}{R_x^2}\,\dfrac{L_{SD}L_D}{L_S} = \dfrac{1}{2}\left(\dfrac{R_E}{R_x}\right)^2 . \tag{98}$$

Equation (86) then takes the form

$$x_1 \pm x = Qf(x_1), \tag{99}$$

where the function $f(x_1)$ is defined by formula (97).

The graph of the dependence $x_1(x)$ for various $Q$ is presented in Fig. 7. It is seen that as distinct from the point-lens case, the second image vanishes for $x > x_{cr} = (\pi/2)Q(3-\alpha)/(2-\alpha)$. Note that the value of $x_1(0)$ may be either greater or smaller than unity.

Let us estimate the quantity $Q$: if a lensed star S is located in the Large Magellanic Cloud (LMC) ($L_S = 50$ kpc) and the lens parameters correspond to the values $M_x = 0.5 M_\odot$, $R_x = 4 \times 10^{14}$ cm indicated in Section 3.4, the most probable $Q$ value (for $L_D = L_S/2$) is $Q \approx 0.04$, i.e., $R_x/R_E \approx 3.5$.

The increase of star brightness $\Omega$ defined by expression (90) now depends on the parameter $Q$. Allowing for Eqn (99), we obtain in our case

$$\Omega_Q = \sum \dfrac{x_1}{x}\,\dfrac{1}{1 - Qf'(x_1)}\, . \tag{100}$$

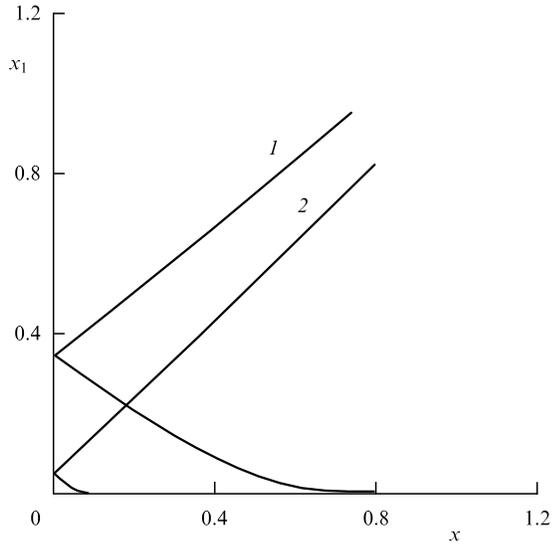

**Figure 7.** Position of images $x_1(x)$ in the case of a non-compact lens for various values of the parameter $Q$: (*1*) $Q = 0.1$, (*2*) $Q = 0.01$.

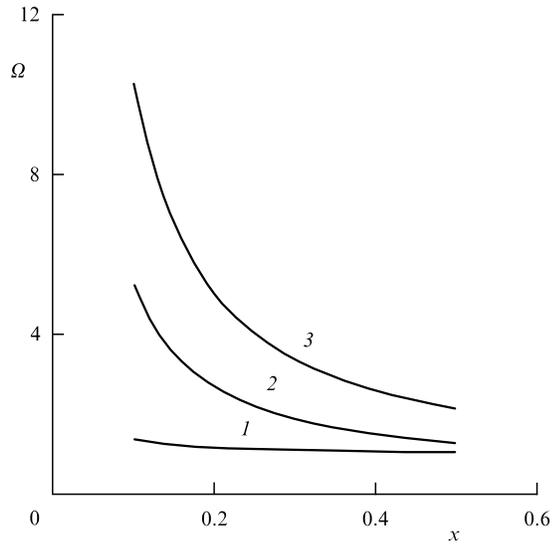

**Figure 8.** Dependence of the amplification coefficient $\Omega(x)$ for various $Q$: (*1*) $Q = 0.01$, (*2*) $Q = 0.1$, (*3*) $Q = 0.4$.

The dependence $\Omega_Q(x)$ for various $Q$ values is presented in Fig. 8. As in the point-lens case, this is a decreasing function going to infinity as $x \to 0$.

The time-dependent variation of parameter $x$ due to the relative motion of the lensing object is described by the formula

$$x = \sqrt{x_{\min}^2 + \left(\frac{v_\perp t}{R_x}\right)^2},$$

where $x_{\min} = \theta_{\min} L_D / R_x$, $\theta_{\min}$ as before is the incidence parameter. With lens motion, the star brightness changes as follows

$$A(t) = \Omega_Q(x(t)).$$

Figure 9 presents the graphs of the dependence $A(t)$ (only for $t > 0$) for various values of the constant $Q$ (or $R_x/R_E$). The time scale $\tau = R_x/v_\perp$ corresponds to $v_\perp = 200$ km s$^{-1}$, $R_x = 4 \times 10^{14}$ cm. For comparison, the $A(t)$ graphs are given for the case of a point source with the same values of $M$ and $L_D$ and the same impact angles $\theta_{\min}$.

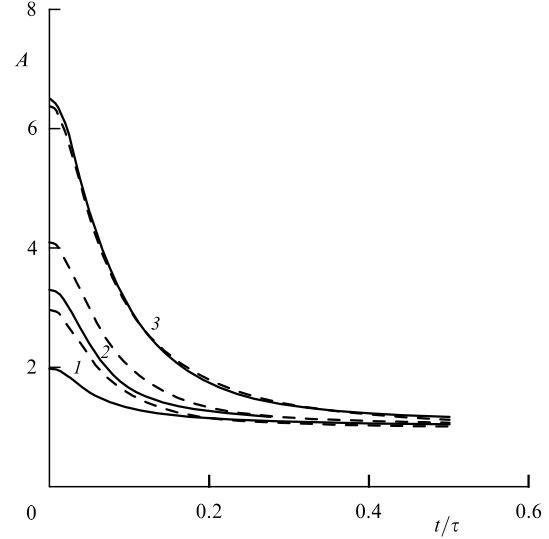

**Figure 9.** Light curve $A(t)$ for various values of $Q$ (the dashed line is a point-like lens): (*1*) $Q = 0.01$, (*2*) $Q = 0.02$, (*3*) $Q = 0.05$. Time is in units $\tau = 2 \times 10^7$ s ($R_x = 4 \times 10^{14}$ cm); $x_{\min} = 0.05$.

One can see that the difference between the curves for $Q \leqslant 0.01$ is not large. It is however of importance that in the case of a non-compact body the shape of the brightness (light) curve is determined not by two but three parameters. As such parameters one can choose the brightness at maximum $A_{\max}$, the lensing duration $\hat{t} = 2R_E/v_\perp$ and $Q = (R_E/R_x)^2/2$. As the scale of the lensing object $R_x$ increases, i.e., $Q$ decreases, the difference between the light curves created by compact and non-compact lenses increases. Their detailed comparison is carried out in the section to follow.

### 5.3 Comparison of light curves created by compact and non-compact lenses

The exact form of the light curve of a compact object is determined by the following parameters: the mass $M_x$ of the lensing body, the distance $L_D$ to this body, the relative velocity $v_\perp$, the incidence parameter $\theta_{\min}$, and the distance $L_S$ to a lensed star. For an NO, there is also a body scale $R_x$. If all of these quantities were known, the difference between the light curves for a compact and a non-compact body as shown in Fig. 9 would be small, but still sufficient to be detected in observations.

The difficulty of the problem is that we do not actually know these parameters. The theoretical light curve is typically chosen so as to attain the best possible agreement with observations. Thus, to determine whether the lensing body is compact or non-compact, we have to compare the closest lensing curves and not those corresponding to similar values of parameters. Let us find out how the difference between these curves can be characterized.

The problem is formulated as follows. The light curve $A_c(t)$ of a compact body with a given amplitude $A_0$ and a characteristic duration $\hat{t}_c$ is defined. In the case of a non-compact lens, the light curve is characterized by three parameters, namely, the amplitude $A_{n0}$, duration $\hat{t}_n$, and size of the lensing body $R_x/R_E$ [or the parameter $Q$ (98)].

Our task is that for a given ratio $R_x/R_E$ the amplitude $A_{n0}$ and the width $\hat{t}_n$ be so chosen that the difference between the compact $A_c(t)$ and non-compact $A_{nc}(t)$ curves should be minimal.

To this end, we shall construct a functional

$$J = \int \left[ A_{nc}(t) - A_c(t) \right]^2 \frac{dt}{\hat{t}_c} \,. \tag{101}$$

This is a dimensionless quantity which we shall just choose as characteristic of the distinction between the compact and non-compact light curves. Now varying arbitrarily the amplitude $A_{n0}$ and the characteristic time $\hat{t}_n$ of the non-compact light curve, we shall find the minimum $J$ for a given $A_c$ and $Q$. This minimum will describe the difference between the closest light curves for a non-compact and a compact lens.

The corresponding closest light curves for different values of the amplitude $A_{c0}$ are shown in Fig. 10 for $Q = 0.01$. One can see that although the size $R_x$ of the NO is much larger than the Einstein radius $R_x/R_E = (2Q)^{-1/2} \simeq 7.14$, the distinction between the light curves is not large. The integral $J$ is only $0.02-0.03$. It increases very slowly with increasing amplitude.

As $Q$ increases, the magnitude of the divergence $J$ grows, although for $R_x/R_E = 10$ and even somewhere higher it still remains insignificant and lies within the present-day experimental error (see Section 6).

We point out that as can be seen from Fig. 10, the most substantial difference is observed on the wings of the light curves, i.e., in the region $|t - t_{min}| > \hat{t}_c$. In order to understand this phenomenon, we shall consider the asymptotics of the light curves in this region. The condition $|t - t_{min}| \gg \hat{t}_c$ is equivalent to $u(t) = \theta(t)/\theta_0 \gg 1$ or $R \gg R_E$. For a point-like (compact) lens we then obtain from Eqn (92)

$$A_c(t) - 1 \simeq \frac{2}{u^4}\,, \quad u(t) = \frac{R}{R_E} \simeq \frac{|t - t_{min}|}{\hat{t}_c/2}\,, \quad R \gg R_E\,. \tag{102}$$

For a non-compact lens, on the wings of brightness for $R \gg R_E$ (102) the region of intermediate asymptotics occurs first of all

$$R_E \ll R \ll R_x \,. \tag{103}$$

The light curve in this region can be calculated assuming

$$x = \frac{R}{R_x} \ll 1\,, \quad Q = \frac{1}{2}\left(\frac{R_E}{R_x}\right)^2 \ll 1\,, \quad \frac{R_E}{R} = \frac{2Q}{x^2} \ll 1\,. \tag{104}$$

In view of the fact that the function $f(\xi)$ is limited, we find from Eqn (99) the position of the image $x_1^+ = x + Qf(x) + O(Q^2)$ which is unique in the region under consideration.

Taking account of the asymptotics of the function $f(\xi)$, we obtain from Eqn (100), as $\xi \to 0$,

$$A_{nc}(t) - 1 \simeq \frac{3-\alpha}{2-\alpha}\frac{\pi}{2}\frac{Q}{x} + O(Qx^{1-\alpha}) \sim \frac{\sqrt{Q}}{u}\,. \tag{105}$$

Thus, in the case of a non-compact lens we have the dependence $A \sim u^{-1}$ instead of $A \sim u^{-4}$ in the 'transition' region (103). This difference is responsible for the appreciably more slowly decreasing light curve wings which are shown in Fig. 10.

Further on, however, the relation between the compact and non-compact light curves changes. For $R \gg R_x$, in a non-compact lens, as is seen from Eqns (97), (99), one image coincides with that of the point lens and the other image is absent. It follows from Eqn (100) that the difference of the amplification coefficient from unity for non-compact lens is half that for a compact lens:

$$A_{nc}(t) - 1 \simeq \frac{4Q^2}{x^4} = \frac{1}{u^4}\,, \quad u(t) = \frac{R}{R_E} \gg 1\,, \quad x = \frac{R}{R_x} \gg 1\,. \tag{106}$$

Therefore, as $t \to \infty$, the light curve corresponding to a non-compact lens runs lower.

### 5.4 Influence of the baryonic core on the lensing curve for a non-compact object

We have already considered the simplest model of density distribution for an NO, which is characterized by a single parameter, namely, the body radius $R_x$. As mentioned in Section 4.2, for the small-scale structure this model is somewhat modified, and the cut-off parameter $r_c$ (56) appears. Furthermore, a considerable influence on the light curve might, in principle, also be exerted by a baryonic body located in the center of an NO. So, let us therefore consider a model with a density distribution (58) of non-baryonic matter

$$\rho = \begin{cases} kR_1^{-\alpha}, & r < R_1; \\ kr^{-\alpha}, & R_1 < r < R_x; \\ 0, & R_x < r \end{cases} \quad (107)$$

and suppose in addition that in the center is a baryonic body which may be thought of as point-like. The gravitational potential $\Phi(r)$ will then assume the form

$$\Phi(r) = -\frac{GM}{R_x} \begin{cases} \frac{3-\alpha}{2-\alpha} + \frac{\lambda}{\beta}\left[\frac{R_1}{r} - \frac{3-\alpha}{2\alpha}\left(\frac{r}{R_1}\right)^2 \right. \\ \left. \qquad -\frac{3}{2}\frac{3-\alpha}{2-\alpha}\right], & r < R_1; \\ \frac{3-\alpha}{2-\alpha} - \frac{1}{2-\alpha}\left(\frac{r}{R_x}\right)^{2-\alpha}, & R_1 < r < R_x; \\ \frac{R_x}{r}, & R_x < r. \end{cases}$$

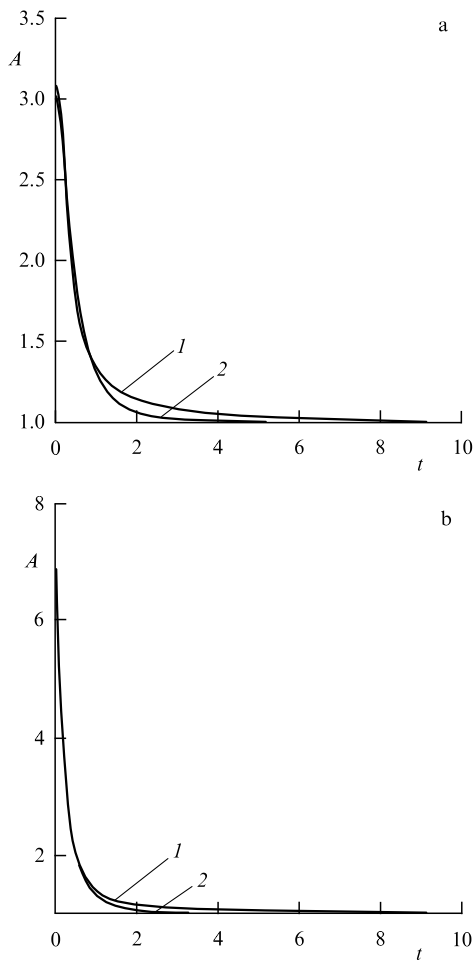

**Figure 10.** 'Non-compact' curves *1* closest to the given 'compact' curves *2*; $Q = 0.01$; amplitudes of 'compact' curves $A_{c0}$: (a) 3, (b) 7. The time is normalized to $\hat{t}_c$.

Here $M$ is the total mass of the object and $\lambda$ is the fraction of baryonic mass. The ratio $\beta = R_1/R_x$ determining the cut-off radius $R_1$ is chosen so that the coefficient $k$ in (107) remains the same as in Eqn (94), i.e., the mass 'deficit' of the non-baryonic matter in the center is equal to the mass of the baryonic core. Then $\lambda = (\alpha/3)\beta^{3-\alpha}$. Further calculations are completely analogous to Eqns (96)–(100) with the only difference that here, as in the case of a point (Schwarzschild) lens, two images are always present.

The corresponding light curves which are particularly close to the Schwarzschild curve of a given amplitude are represented in Fig. 11. The value of the parameter $\lambda$ is chosen to be 5% and the cut-off radius corresponds to $r_c$ (57). One can see that the light curves both with and without allowance for the baryonic core are almost coincident. A substantial difference from the case of a compact body is pronounced, as before, on the wings of the light curve.

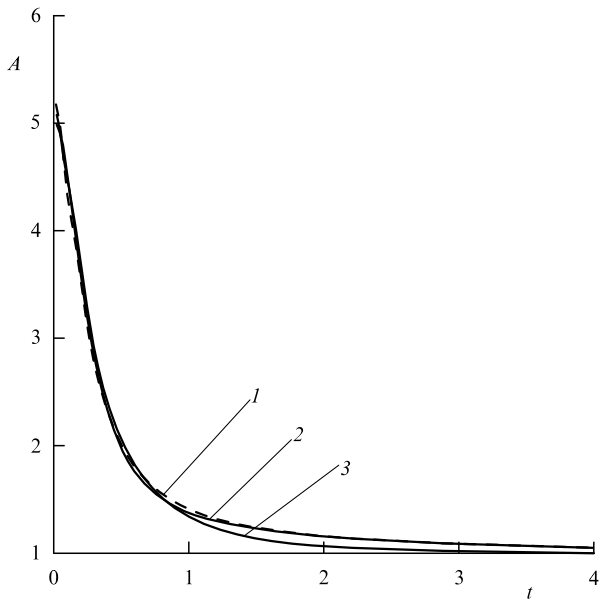

**Figure 11.** Light curves due to a non-compact lens both with and without baryonic core which are closest to the given Schwarzschild curve: (*1*) NO with a baryonic core (the dashed line), (*2*) NO without a baryonic core, (*3*) a compact object.

### 5.5 Optical depth
We have considered individual characteristics of the microlensing curves. An important role is played by the statistical characteristics, primarily the probability of observing the microlensing effect. This probability is naturally characterized by the 'optical depth' $\tau$.

The quantity $\tau$ characterizes the probability that the light of a given star experiences lensing at a given moment of time. This probability is defined by the relation

$$\tau = \int_0^{L_S} \sum_M n_M(l)\sigma_{Ml}\, dl. \quad (108)$$

Here $n_M(l)$ is the concentration of lensing bodies of mass $M$ in space, $\sigma_{Ml}$ is the effective microlensing cross-section, and $L_S$ is the distance to the lensed star. It is natural to set $\sigma_{Ml} = \pi R_E^2$, where $R_E$ is the Einstein radius (88). Then [17]

$$\tau = \frac{4\pi G}{c^2} \int_0^{L_S} \frac{l(L_S - l)}{L_S} \sum_M Mn_M(l)\, dl. \quad (109)$$

This shows that the optical depth is, in fact, independent of the mass $M$ and is only determined by the distribution of the total density $\rho_d(l) = \sum_M Mn_M(l)$ of the dark matter contained in these objects, in the space between the observer and the lensed star.

## 6. Microlensing — results of observations and comparison with the theory

### 6.1 Microlensing of objects in the halo and the central part (bulge) of the Galaxy
The probability of observing the stellar microlensing effect described by the 'optical depth' $\tau$ is very small in the real conditions of the Galaxy. The quantity $\tau$ can be readily evaluated. Indeed, as follows from Eqn (109),

$$\tau \sim \frac{G\rho L_S^2}{c^2} \sim \frac{v^2}{c^2} \sim 5 \times 10^{-7}. \quad (110)$$

Here $\rho$ is the mean dark-matter density in a region with a characteristic scale $L_S$ and $v$ is the characteristic velocity of dark-matter motion in this region. In the Galaxy, as is well-known, the velocity is $v \sim 200$ km s$^{-1}$, which implies the estimate (110).

The characteristic lensing time, $\hat{t} \sim R_E/v$, by an object of mass $(0.1-1)M_\odot$ located at a distance $L_D \simeq 10$ kpc makes up a value of the order of a month or a year. Therefore, the effect of microlensing of a star located in the Galactic halo can be seen with a rather high probability $p \sim 1$ just once within the time $\hat{t}/\tau$, i.e., a time of the order of one million years. From this it is clear that observation of stellar microlensing can actually be carried out only by simultaneously tracing the radiation intensity variation of not less than a million stars. Such a possibility was realized using special modern technical devices — large CCD matrices with 2048 × 2048 cells. These studies were started by the MACHO [14], EROS [41], and OGLE [45] groups.

In studies of the halo, the sources of lensing light are stars from the Large and Small Magellanic Clouds and also from the galaxy M31 (Andromeda). Up to the present time microlensing could only be observed for LMC stars. For this purpose, using a 127-cm telescope on Stromlo mountain in Australia, the MACHO group is conducting continuous observations of the intensity of 8.5 million stars.

Observations of the central part of the Galaxy are being carried out in the region free of absorption by dust — the so-called 'Baade window' (galactic coordinates $l = 2.5°$, $b = -3.6°$). The MACHO observes 12.6 million stars and the OGLE group observes nearly the same number.

It should be specially pointed out that this is the first attempt to observe the intensity of such a huge number of stars simultaneously, and the process itself must provide invaluable material for the investigation of star variability. However, such a large amount of data entails considerable difficulties in revealing microlensing cases. Indeed, as is clear from relation (110), only one star out of several million shows a noticeable change in brightness due to microlensing. At the same time, the number of stars that simply possess non-stationary radiation is three orders of magnitude greater. For this reason, the authors of the experimental studies [14, 41] developed special methods of processing the observational data and found criteria for identifying microlensing. An important role among these criteria is played by the achromatism of the gravitational lensing process. Therefore observations are carried out simultaneously in two chromatic ranges — blue ($\lambda \approx 4500-6000$ Å) and red ($\lambda \approx 6300-7300$ Å). Other important criteria of selection are connected with the requirement that the observational data should be in agreement with the theoretical curve (92) (see Fig. 6) that describes the intensity variation with time. It is natural that microlensing is always assumed here to be determined only by the action of compact invisible objects (the name of the MACHO group just means 'massive compact halo objects').

We note that the shape of the theoretical curve of lensing by a compact object, as we have seen above (92), is defined uniquely by two parameters (for instance, amplitude and duration), whereas the number of unknown quantities determining these parameters and, accordingly, affecting the shape of the curve is much larger. Hence, to estimate the mass of a lensing object one has to make certain assumptions about the values of other parameters. So, in observations in the halo, the most probable values $v_\perp = 200$ km s$^{-1}$ and $L_D = 10$ kpc are usually used, which makes it possible to determine the values of the object's mass $M$ and the characteristic scale $R_E$ from the results of observations.

### 6.2 Results of observations in the halo
Observations of microlensing on objects in the halo are carried out mainly by the MACHO group [15, 16] (the EROS group observed two objects in 1993 [41]). Examples of MACHO observations are given in Fig. 12. During two years of observations they discovered eight cases of microlensing. In Fig. 12 one can see that only observations whose results agreed sufficiently well with the theoretical curve of lensing by a compact body (92) were selected. This agreement is however not sufficiently perfect to state that it is compact objects that are revealed in the observations (see Section 6.3).

The characteristic lensing time $\hat{t}$ for observed objects differs a little and lies between 34 and 145 days. This corresponds to a fairly large mass $M_x$ of objects. The masses of the objects observed almost reach the mass of the Sun $M_\odot$. The number of microlensing events is also very large. For example, as calculations show, if microlensing took place only on the known invisible star component of the Galaxy, a substantially (over an order of magnitude) smaller amount of events would occur. The same refers to LMC stars. Thus, observational data show quite definitely that microlensing is caused by objects belonging to the halo.

The characteristic optical depth $\tau$, determined on the basis of observational data, appears to be equal to

$$\tau = \left(2.9^{+1.4}_{-0.9}\right) \times 10^{-7}. \tag{111}$$

The statistical processing of the results of observations is presented in Fig. 13. Here $M_x$ is the mass of lensed objects and $f$ is their fraction in the halo as compared to the total dark-matter mass determined from the rotation curve. One can see that the mean mass of observed objects is rather large and, which is even more important, they make up more than half of the total dark matter in the halo.

It is particularly noteworthy that it takes a very long time to single out large-scale objects from observational data. It is therefore not surprising that it is not only the number but also the mean mass of the objects revealed by the MACHO group that rises gradually with increasing observations (compare [15] and [16]). It is not excluded that this process will go longer. If one assumes, in line with the authors of the experimental papers, that the objects observed in microlensing are ordinary stars (brown and white dwarfs), one cannot but notice that the results of observations lead to increasing contradictions with the conventional opinion, based on the data on the nucleosynthesis and the origin of galaxies (see Introduction and [46]), that the dark matter is of non-baryonic origin. Moreover, it becomes difficult to find agreement between the interpretation of microlensing data from observation of brown and white dwarf stars in the halo and the results of direct optical observations on the Hubble telescope [18].

It should be noted that observational data do not satisfy the criteria of exact agreement with the theory of microlensing on compact objects. The error $\chi^2$ describing the deviation of observational data from the theoretical curve exceeds 3–5 and sometimes even 10–20 times the statistically admissible quantity $\sigma$. The authors of Ref. [16] explain this divergence by various types of errors in observational data, by microlensing of giant stars, and so on. But it is undoubted that *there is no experimental proof of the fact that the objects in the halo, observed in microlensing, are compact bodies.*

In this connection it is of particular interest to clarify the results of comparison of observational data with the theory of microlensing on non-compact (non-baryonic) objects.

### 6.3 Comparison of the data of observations in the halo with the theory of microlensing on non-compact objects
It was shown by the analysis presented in Section 5.3 that for not very large sizes $R_x$ of a non-compact body,

$$\frac{R_x}{R_E} \leqslant 10, \tag{112}$$

the microlensing curves look outwardly like the curves for compact bodies. The most thoroughly investigated event was reported in detail by the MACHO group in Ref. [47] (Fig. 14). We used these data, which are available to us, to compare observational results with the theory [13, 19, 48].

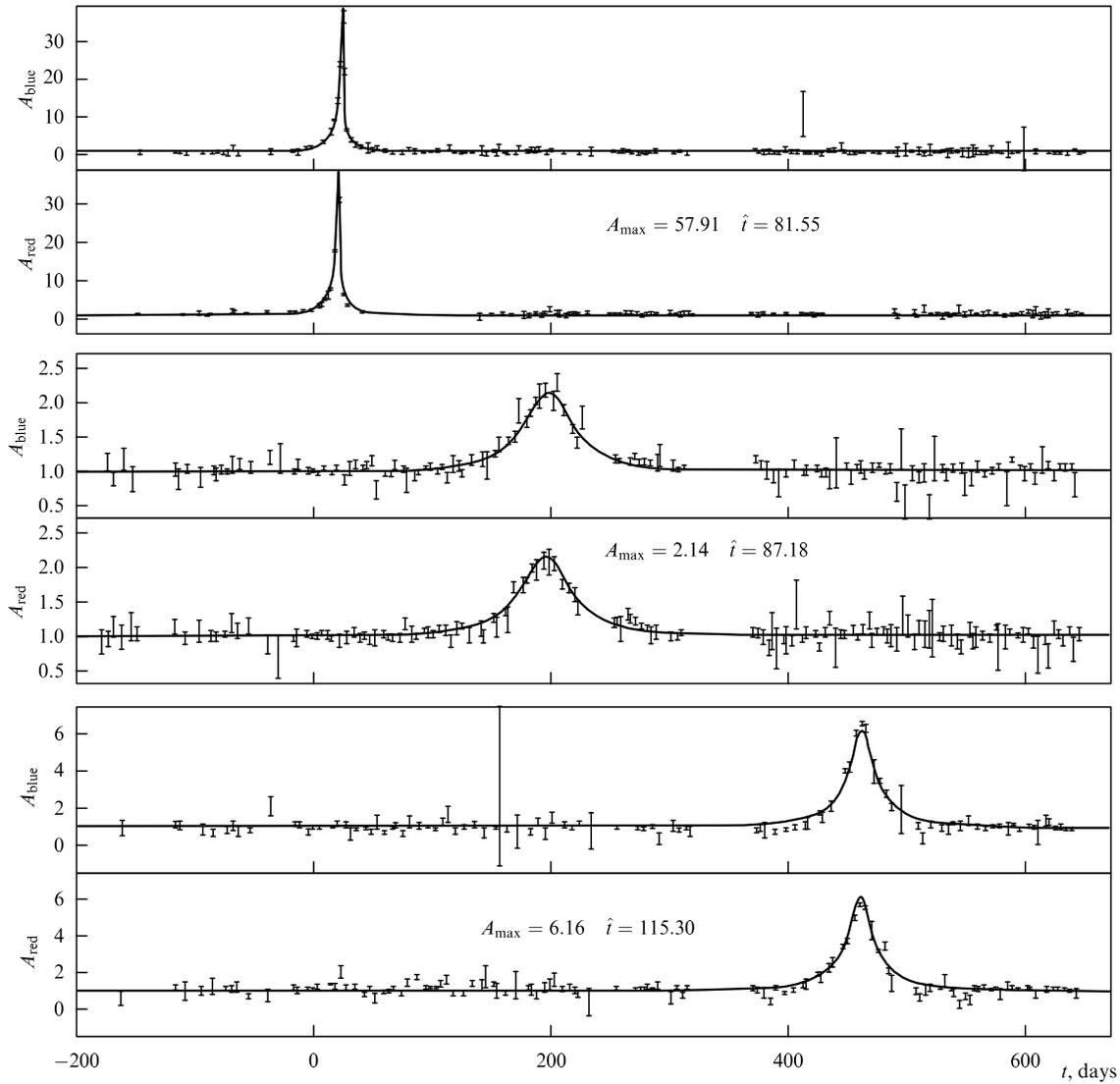

**Figure 12.** Examples of microlensing curves according to MACHO observations. For each event, the data are given in two colors (red and blue). The points stand for the results of observations, the curve represents the theoretical interpretation.

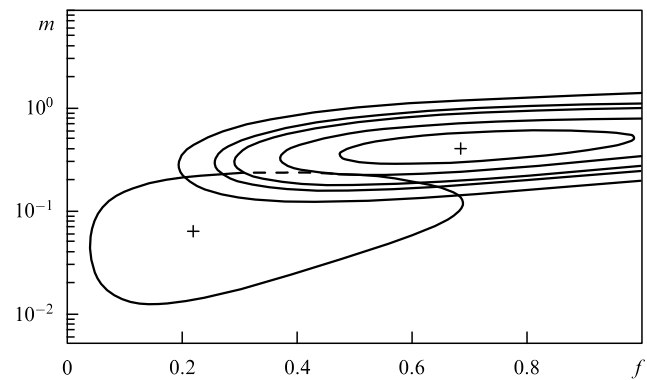

**Figure 13.** Statistical processing of the results of observations carried out by the MACHO group: $f$ is the fraction of the total mass of the halo contained in the objects; $m$ is objects' mass in units of solar mass $M_\odot$. The most probable value is marked by +, the contours correspond to the probabilities 34%, 68%, 90%, 95%, 99%. Given separately are the results of observations before 1995 [15].

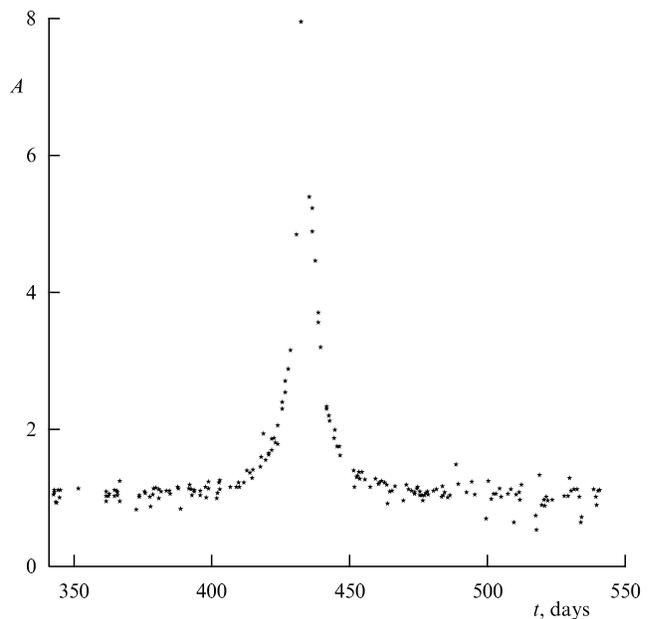

**Figure 14.** Data of observations [47].

The comparison is made using the $\chi^2$ criterion

$$\chi^2 = \sum_i \frac{[A(t_i) - A_{\text{obs},i}]^2}{\sigma_{\text{obs},i}^2} \,. \tag{113}$$

Here $t_i$ is the instant of time when an observation took place, $A_{\text{obs},i}$ and $\sigma_{\text{obs},i}$ are an observed amplitude and an experimental error, and $A(t)$ is the theoretical curve. The latter depends on several parameters. In the case of a compact body these parameters are the position of the maximum of $t_{\min}$, the amplitude at the maximum $A_0$ and the characteristic time $\hat{t}$ (see Section 5.1). Also of great importance is a sufficiently exact determination of a constant level of star brightness long before or long after the lensing because it is to this level that the observational data are normalized.

In the case of a NO there is at least one more parameter characterizing the size of the body $R_x$; it is natural to use it in the dimensionless form $R_x/R_E$ or $Q$ (98).

The parameters are so chosen that the functional $\chi^2$ reaches its minimum, and the bulk of the observational data is used. In Refs [13, 48], the following values of the parameters are given:

(1) for a compact body:

$$t_0 = 433.604, \quad A_0 = 7.51, \quad \hat{t} = 35.5, \quad \chi^2_{\min} \approx 309; \tag{114}$$

(2) for a non-compact body:

$$t_0 = 433.73, \quad A_0 = 7.72, \quad \hat{t} = 35.3, \quad \chi^2_{\min} \approx 304. \tag{115}$$

Note that in the latter case the minimum of $\chi^2$ for $Q = 0.03$, i.e., $R_x/R_E = 4.22$, is quite clearly pronounced. The data of 253 observational points $t_i$ (in blue) were employed in the calculation. Hence, the error is $\sigma = \sqrt{2N} \approx 23$, and therefore a value $\chi^2_{\min}$ within the limits $240 \leqslant \chi^2_{\min} \leqslant 286$, i.e., within the limits of $1\sigma$, would signify a coincidence between the theoretical curve and observations. The divergence, as seen from the values of the parameters (114) and (115), although existent, is not large: for a compact body it is $1.91\sigma$ and for a non-compact one — $1.78\sigma$. Obviously, on the basis of these data it is impossible to make a definite statement of whether the observed object is compact or non-compact, although there is some evidence in favor of a non-compact body. One should stress once again the particularly important role of determining of a constant level of star brightness $\hat{I}$ — although we used for this purpose the data of nearly two years of observations (with an effective lensing duration of $\hat{t} \sim 35$ days), the error in $\hat{I}$ measurements remained large enough to affect the determination of the quantity $\chi^2_{\min}$.

Comparison of the theoretical curves for the values obtained for the parameters (114), (115) with observational data is presented in Fig. 15. Their generally good agreement is obvious. The substantial difference of curves for non-compact and compact objects that occurs on the wings (see Fig. 15) lies within experimental error in our case. The mass of the lensing object evaluated in Ref. [13, 48] in this case is equal to $M_x \approx 0.05 M_\odot$. For the parameters (114) and (115) it then follows that its size is

$$R_x \approx 4.22 R_E \approx 1.3 \times 10^{14} \text{ cm} \,. \tag{116}$$

This value is in agreement with the theoretical estimates (41).

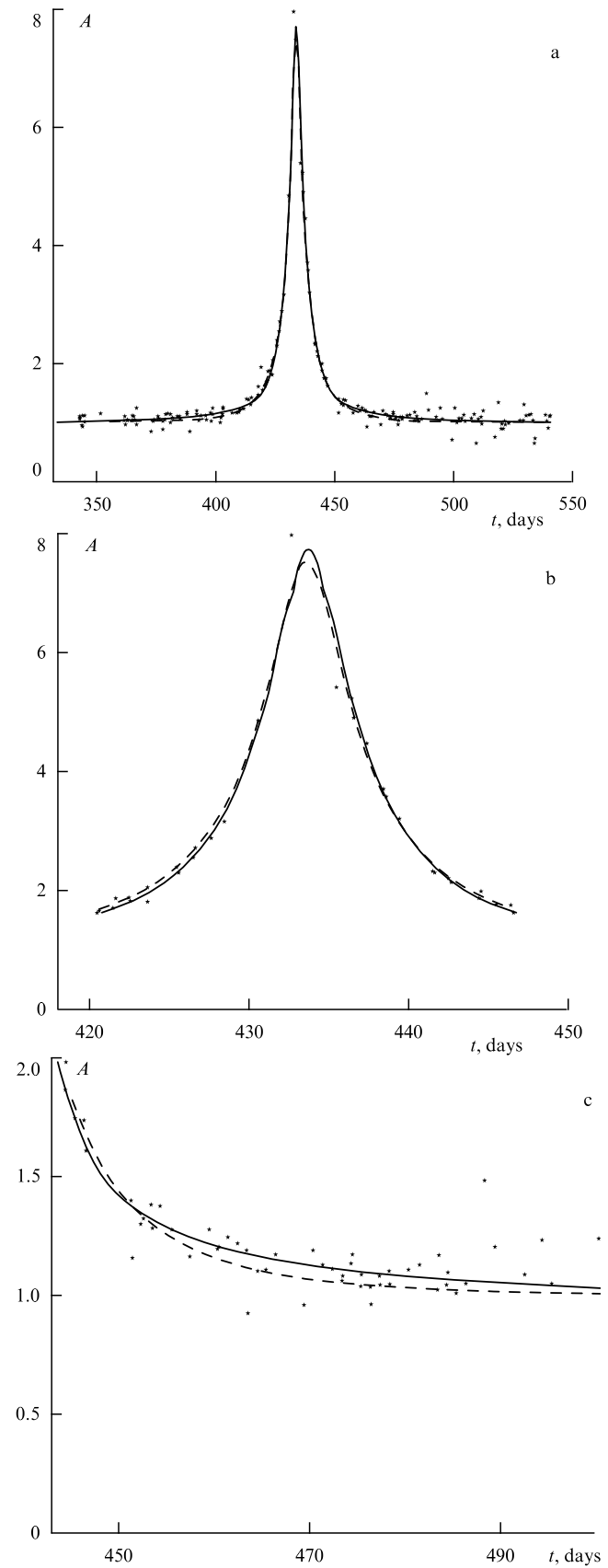

**Figure 15.** Comparison of the theoretical curve (the result of processing) with observations. The solid curve corresponds to a non-compact lens and the dashed line to a point lens; (a) the whole curve, (b) the central part, (c) the right wing.

The comparison thus demonstrated that the data considered do not make it possible to state that an object observed in the halo is non-compact and, therefore, non-baryonic. At the same time there are no grounds to think of it as a compact baryonic body. This question needs further investigation.

### 6.4 Results of observations in the central region of the Galaxy and a comparison with theory

Observations in the central region of the Galaxy reveal a much greater possibility of microlensing: over one year of measurements, the MACHO group observed 45 events [49] and nearly the same number was reported by the OGLE group [50]. Figure 16 gives examples of the results of observations. An important peculiarity of microlensing in the bulge is that it was realized for both ordinary stars and the subgroup of giant stars (about 1.3 million). The distances to these stars and their luminosities are well known. This allowed a more precise estimate of the optical depth $\tau$ which proved to be fairly large:

$$\tau = \left(3.9^{+1.8}_{-1.2}\right) \times 10^{-6},$$

which is a much larger quantity than that given by the calculations within standard theoretical models of star distribution in the bulge. One may thus hope that the use of microlensing will provide an insight into the structure of the central part of the Galaxy.

In the microlensing of giant stars occur deviations of the light curve from the curve of ordinary 'point-star' lensing (Section 6.1). These deviations were revealed in recent observational data [51].

Note that in the central part of the Galaxy, the mass of the baryonic component in the center of an NO may be substantially higher than that in the halo (see Section 4.5).

Microlensing realized in the central part of the Galaxy by invisible stars (i.e., baryonic objects) is, of course, much more probable than in the halo. However, in this region there may also exist non-compact baryonic objects with lifetimes, as shown in Section 3.5, much greater than the lifetime of the Universe (46). That is why it is of great interest to compare observational data with the microlensing theory on NO in this region, too. Such a comparison was performed in the paper [52] where the microlensing data obtained by the OGLE group were analysed. Out of the six curves examined, in three cases the microlensing data appeared to be considerably closer to the theory of a compact body. In one of the cases presented in Fig. 17, the picture however is the opposite: the microlensing theory for NO provides a much better description of the observational data. For a NO, the error $\Delta = \chi^2_{\min} - N$ (where $N$ is the number of degrees of freedom) is

$$\Delta \approx 2.1\sigma, \quad \sigma = \sqrt{2N}, \qquad (117)$$

whereas for a compact object this value is

$$\Delta \approx 6.2\sigma.$$

So we see that comparison of the theory with observational data in the central part of the Galaxy also gives indications of the possible existence of NO.

### 7. Neutralino stars

The properties of NO considered up-to now are independent of the nature of dark-matter particles. It is presently unclear what particles constitute dark matter, there existing various hypothetical candidates: neutralinos, heavy neutrinos, axions, and strings. Neutralinos and heavy neutrinos are the

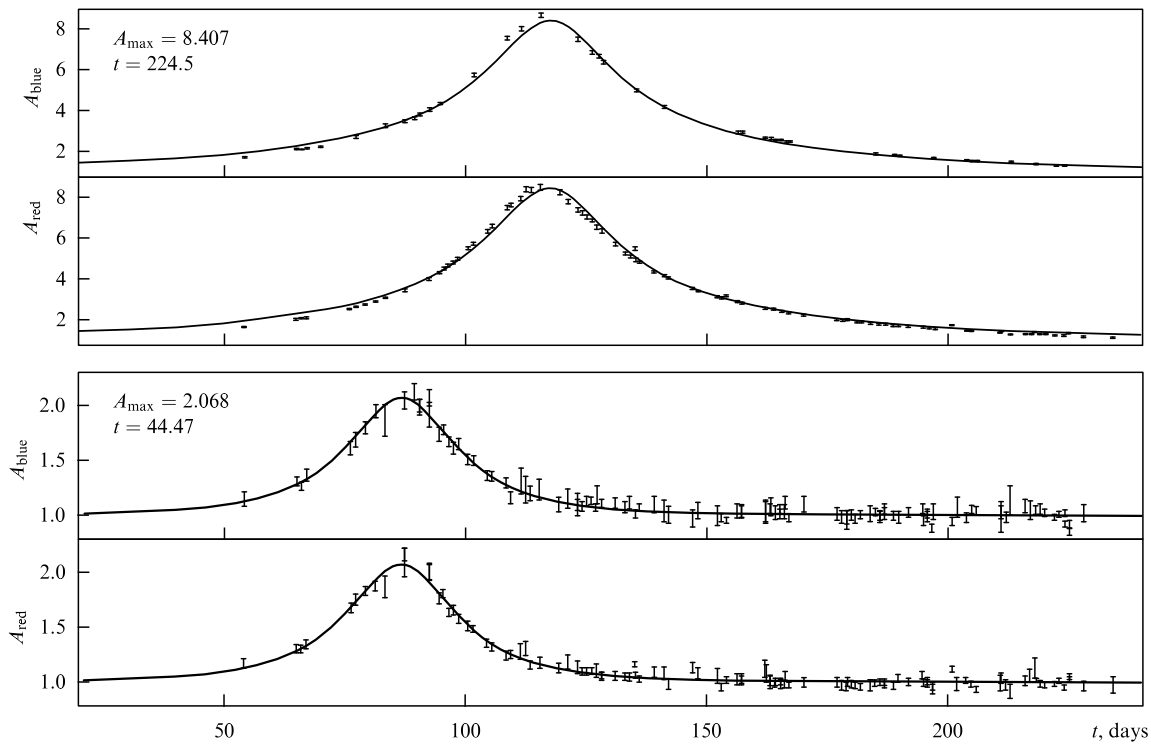

**Figure 16.** Examples of the results of observations carried out by the MACHO group in the central region of the Galaxy [49].

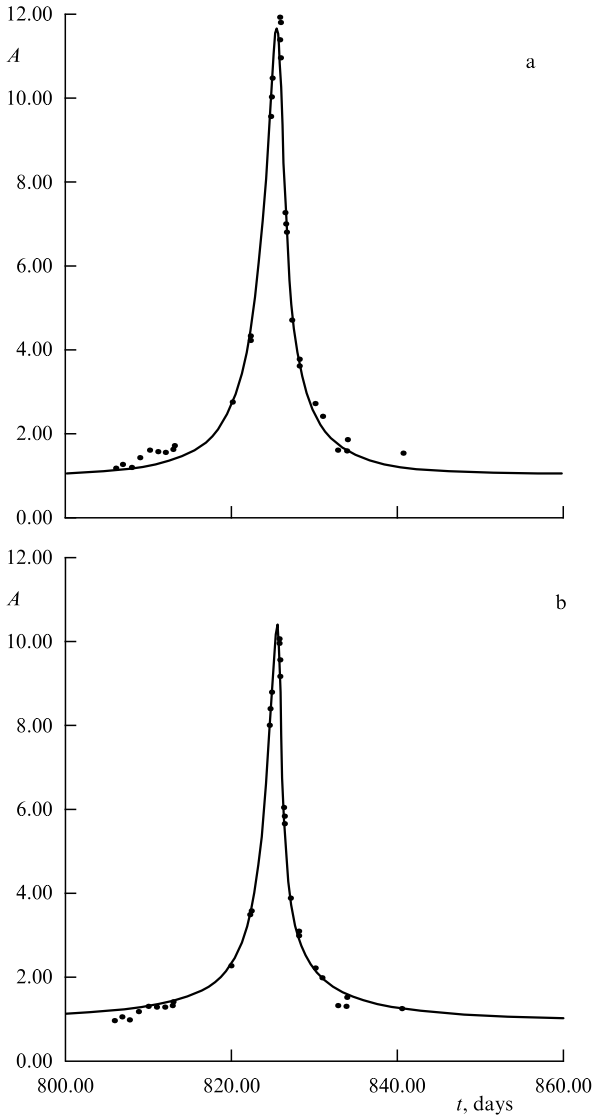

Figure 17. Processed data of OGLE's observations: the 'non-compact' case [52]. Theoretical curves: (a) a point lens, (b) a non-compact lens.

Majorana particles, and therefore the crucial point of interaction between these particles is annihilation. If NO consist of such particles, they should dissipate partially owing to annihilation. Such objects will be called Neutralino Stars (NeS) [12] and the constituent particles — 'neutralinos'. We shall stress that the word 'neutralino' is used here for all CDM particles annihilating in collisions.

Neutralino annihilation leads to intense gamma-quantum emission, baryonic body heating, etc. It should be emphasized that the energy released in these processes was gained in CDM in the course of dark-matter particle decoupling. The possibility of discovering these particles by their annihilation has been discussed previously (see Refs [53, 34]). In these papers, however, the density distribution of CDM particles was always assumed to be either smooth or gradually increasing towards the center of the Galaxy. The possibility of the existence of NeS suggests quite a new situation connected with an essentially non-uniform density distribution of neutralinos in the halo and in the Galaxy, and their strong compression in NeS would, therefore, cause a substantial strengthening of the annihilation processes. On the one hand, this is promising for what concerns observation of non-compact dark-matter objects (NeS in the given case) and on the other hand, this may serve as a source of important information on CDM particles [54].

### 7.1 Extragalactic diffusive gamma-radiation

Let us consider gamma-ray fluxes due to neutralino annihilation in NeS. A diffusive gamma-ray flux $I_\gamma$ is determined by two sources: extragalactic gamma-radiation $I_{1\gamma}$ coming from NeS and a radiation flux $I_{2\gamma}$ from NeS located in the halo of our Galaxy.

$$I_\gamma = I_{1\gamma} + I_{2\gamma} \,. \tag{118}$$

Extragalactic radiation is determined by complete energy loss in the Universe due to annihilation in NeS:

$$I_{1\gamma} = \frac{c}{4\pi\sqrt{3}} \, t_0 \dot{\varepsilon}_h \alpha_\gamma \ (\text{cm}^{-2} \ \text{s}^{-1} \ \text{sr}^{-1}) \,. \tag{119}$$

Here $I_{1\gamma}$ is the number of gamma-quanta passing through 1 cm² per second per steradian, $c/\sqrt{3}$ is the photon velocity averaged over angles, $\dot{\varepsilon}_h$ are the mean annihilation energy losses per second in 1 cm³, $\alpha_\gamma$ is the coefficient of energy conversion into gamma-quanta, and $t_0$ is the lifetime of the Universe before the red shift $z \approx 1$ (a cut-off for $z \approx 1$ is due to a red shift for a sufficiently rapid fall of the gamma-radiation spectrum with increasing energy). The energy loss is specified by the relation

$$\dot{\varepsilon}_h = c^2 \rho_c \Omega f \tau^{-1} \,, \tag{120}$$

where $\rho_c$ is critical density in the Universe,

$$\rho_c = \frac{3H^2}{8\pi G} \,, \tag{121}$$

$H$ is the Hubble constant, $\Omega$ is the ratio of the mean density of matter to the critical density, $f$ is the fraction of dark matter in the Universe contained in NeS, and $\tau$ is the neutralino lifetime in an NeS:

$$\tau = \frac{N_x}{|\dot{N}_x|} \,. \tag{122}$$

Here $N_x$ is the total number of neutralinos in an NeS,

$$N_x = \frac{M_x}{m_x} \,,$$

$M_x$ is the NeS mass, $m_x$ is the neutralino mass, $\dot{N}_x$ is the neutralino annihilation rate,

$$\dot{N}_x = -4\pi \langle \sigma v \rangle \int n_x^2(r) r^2 \, dr \,, \tag{123}$$

$n_x(r)$ is the neutralino density in an NeS

$$n_x(r) = \frac{\rho}{m_x} \,, \tag{124}$$

and $\langle \sigma v \rangle$ is the averaged product of annihilation cross-section $\sigma$ with particle velocity $v$. From Eqns (123), (124) and (58) it follows that

$$|\dot{N}_x| = \frac{(3-\alpha)^2}{4\pi(2\alpha-3)} \left(\frac{M_x}{m_x}\right)^2 \frac{\langle \sigma v \rangle}{R_x^3} \left[\left(\frac{R_x}{r_c}\right)^{2\alpha-3} - 1\right] \tag{125}$$

and

$$\tau = \frac{4\pi(2\alpha - 3)}{(3 - \alpha)^2} \frac{m_x}{M_x} \frac{R_x^3}{\langle \sigma v \rangle} \left[ \left( \frac{R_x}{r_c} \right)^{2\alpha - 3} - 1 \right]^{-1} . \quad (126)$$

For the density $\rho(r)$ we have used the relations (58) and (59). Proceeding from Eqns (119)–(122) we find the flux of extragalactic gamma-radiation:

$$I_{1\gamma} = \frac{\sqrt{3}}{32\pi^2} \frac{\alpha_\gamma f \Omega c^3 H^2 t_0}{G\tau} . \quad (127)$$

### 7.2 Diffusive gamma-radiation from the galactic halo

A detected gamma-ray flux from a NeS, which is generated in our Galaxy, depends on the direction of observation:

$$I_{2\gamma}^h(\theta, \phi) = \frac{1}{4\pi} \dot{N}_\gamma \int_0^{R_h} n(\mathbf{r}) \, dr . \quad (128)$$

Here $\dot{N}_\gamma$ is the number of gamma-photons emitted per second by one NeS and $n(\mathbf{r})$ is the density of NeS in the halo and the Galaxy. According to Eqn (43), the NeS density in the halo is

$$n(\mathbf{r}) = \frac{3 - \alpha}{4\pi} \frac{N_s}{R_h^3} \left( \frac{|\mathbf{r} - \mathbf{r}_\odot|}{R_h} \right)^{-\alpha}, \quad \alpha \approx 1.8 , \quad (129)$$

where $\mathbf{r}_\odot$ is the coordinate of the Sun and $N_s$ is the total number of NeS in the halo (42):

$$N_s = f_h \frac{M_{dh}}{M_x} , \quad (130)$$

$M_{dh}$ is the total dark-matter mass in the halo, $R_h$ is the halo size, and $f_h$ is the fraction of the halo's dark matter contained in NeS.

Choosing the coordinate system along the vector $\mathbf{r}_\odot$ we obtain that the flux $I_{2\gamma}$ does not depend on the angle $\phi$. The dependence on the angle $\theta = \arccos(\mathbf{rr}_\odot/rr_\odot)$ is described by the integral

$$J = \int_0^{R_h} R_h^\alpha (r^2 + r_\odot^2 - 2rr_\odot \cos \theta)^{-\alpha} \, dr \approx \frac{R_h^\alpha}{r_\odot^{\alpha - 1}} F(\theta), \quad (131)$$

$$F(\theta) = \int_0^\infty (x^2 - 2x \cos \theta + 1)^{-\alpha} \, dx . \quad (132)$$

In integral (132) it was taken into account that $R_h/r_\odot \gg 1$. A simple approximation for this integral can be obtained if we assume $\alpha = 2$:

$$F(\theta) = \frac{\pi - \theta}{\sin \theta} . \quad (133)$$

Numerical integration of Eqn (132) shows that the approximate expression (133) coincides with (132) to within several percent.

Allowing for

$$\dot{N}_\gamma = \alpha_\gamma m_x c^2 |\dot{N}_x| , \quad (134)$$

we obtain from Eqns (125)–(132)

$$I_{2\gamma}^h = I_{2\gamma}^0 F(\theta) , \quad (135)$$

where $I_{2\gamma}^0$ is the flux $I_{2\gamma}^h$ in the anticenter direction $\theta = \pi$,

$$I_{2\gamma}^0 = \frac{3 - \alpha}{16\pi^2} \alpha_\gamma f \frac{M_{dh} c^2}{R_h^2 \tau} \left( \frac{R_h}{r_\odot} \right)^{\alpha - 1} . \quad (136)$$

Of interest is the relation $p_\gamma$ between the galactic $I_{2\gamma}$ and extragalactic $I_{1\gamma}$ diffuse fluxes. From Eqns (127) and (136) it follows that

$$p_\gamma = \frac{I_{2\gamma}^0}{I_{1\gamma}} = \frac{\sqrt{3}(3 - \alpha)}{4\pi} \left( \frac{f_h}{f} \right) \frac{M_{dh}}{\Omega \rho_c R_h^2 c t_0} \left( \frac{R_h}{r_\odot} \right)^{\alpha - 1} . \quad (137)$$

### 7.3 Diffusive gamma-radiation of the Galaxy

NO, as was mentioned above, are centers of interstellar gas condensation, and therefore the mass of their baryonic core $M_b$ may increase substantially. As shown in Section 4.5, owing to the gravitational effect of the baryonic core, a neutralino star undergoes compression (78) and the neutralino density increases (74), (75).

This leads to an increase of gamma-radiation. Indeed, as follows from (123), the gamma-radiation of a NeS is proportional to the neutralino density squared, and therefore allowing for expressions (75) and (78), for the coefficient of compression we obtain

$$I_{NeS}^G \approx p I_{NeS}^h , \quad p = \varkappa_{dm}^2 = \left[ 1 + 3.6 \left( \frac{M_b}{M_x} \right)^2 \right]^6 , \quad (138)$$

where $I_{NeS}^G$ is the neutralino star radiation flux in the Galaxy, $I_{NeS}^h$ is the NeS radiation flux in the halo, and $M_b$ is the baryonic core mass. One can thus see that the gamma-radiation of NeS in the Galaxy may increase several times or even by an order of magnitude owing to the presence of baryonic bodies of mass $M_b \geqslant 0.3 M_x$. This also causes a notable increase of diffusive radiation $I_{2\gamma}^G$ in the Galaxy, the diffusive gamma-radiation $I_{2\gamma}^G$ being specified, by analogy with (128), by the relation

$$I_{2\gamma}^G(\theta, \phi) = \frac{1}{4\pi} \int_G \frac{n_G(\mathbf{r})}{|\mathbf{r} - \mathbf{r}_\odot|^2} \, dV . \quad (139)$$

Here $n_G(\mathbf{r})$ is the distribution of galactic NeS and the integration in Eqn (139) is carried out over the galactic volume G.

### 7.4 Discrete sources of gamma-radiation

NeS may not only contribute to diffusive radiation, but may also be a discrete source of gamma-radiation. Let us determine the intensity of such a source. The formulae (134) and (125) yield the total number of photons emitted per second by a single NeS. From Eqn (134) it follows that the intensity of gamma-radiation of a single NeS at a distance $r_x$ from the source is

$$I_\gamma = \frac{1}{4\pi} \alpha_\gamma m_x c^2 |\dot{N}_x| r_x^{-2} . \quad (140)$$

Since according to Eqn (43) the density of NeS in the vicinity of the Sun is

$$n_s = \frac{3 - \alpha}{4\pi} \frac{N_s}{R_h^3} \left( \frac{R_h}{r_\odot} \right)^\alpha , \quad (141)$$

with probability of the order of unity one can observe several NeS as sources of gamma-radiation at a characteristic distance

$$r_x \simeq r_0 = \left( \frac{3}{4\pi n_s} \right)^{1/3} \quad (142)$$

with intensities

$$I_{\gamma c} = \frac{(1-\alpha/3)^{2/3}}{4\pi}\,\alpha_\gamma f_{\rm h}^{2/3} c^2\,\frac{M_{\rm dh}^{2/3} M_{\rm x}^{1/3}}{\tau R_{\rm h}^2}\left(\frac{R_{\rm h}}{r_\odot}\right)^{2\alpha/3} p\,. \quad (143)$$

Here $p$ is the parameter of NeS radiation amplification at the expense of a massive baryonic core (138). The quantity $I_{\gamma c}$ thus gives the characteristic fluxes from the nearest NeS that may, in principle, be discovered in observations.

### 7.5 Galaxies as distributed sources of gamma-radiation
When calculating the intensity of gamma-radiation from the nearest galaxy M31 (Andromeda), one can use Eqns (128)–(129) replacing $r_\odot$ by $r_{\odot a}$, which is the coordinate of the Sun in the reference frame fixed to the center of Andromeda. Furthermore, when integrating over d$r$ one should remember that the NeS density in Andromeda is described by the relation

$$n_{\rm a}({\bf r}) = \frac{3-\alpha}{4\pi}\,\frac{N_{\rm s}}{R_{\rm ha}^3}\left(\frac{|{\bf r}-{\bf r}_{\odot a}|}{R_{\rm ha}}\right)^{-\alpha}\Theta\left(1-\frac{|{\bf r}-{\bf r}_{\odot a}|}{R_{\rm ha}}\right). \quad (144)$$

Here $R_{\rm ha}$ is the size of the dark-matter halo, $\Theta$ is the Heaviside function. The presence of the $\Theta$-function changes the limits of the integral (128). Taking into account that the size of Andromeda $R_{\rm ha}$ is smaller than $r_{\odot a}$, from Eqns (128) and (144) we obtain

$$I_\gamma^{\rm A} = \frac{2(3-\alpha)^3}{(4\pi)^3(2\alpha-3)}\,\frac{\alpha_\gamma f c^2 M_{\rm dh}^{\rm A} M_{\rm x} r_{\odot a}^{1-\alpha}}{m_{\rm x} R_{\rm ha}^{3-\alpha}}\,\frac{\langle\sigma v\rangle}{R_{\rm x}^3}$$
$$\times\left[\left(\frac{R_{\rm x}}{r_{\rm c}}\right)^{2\alpha-3}-1\right]K(\theta)\,,$$
$$K(\theta) \approx \frac{1}{|\sin\theta|^{\alpha-1}}\arctan\sqrt{\frac{R_{\rm ha}^2}{r_{\odot a}^2\sin^2\theta}-1}\,, \quad (145)$$

where $\theta$ is the angle measured from the direction to the center of Andromeda. From Eqns (145) it follows that the gamma-radiation from Andromeda is distributed over the characteristic angles $\theta \approx R_{\rm ha}/r_{\odot a} \sim 20°$. Integrating expression (145) over the angles $\phi$ and $\theta$, we obtain the total gamma-ray flux $I^{\rm A}$ from the halo of M31:

$$I^{\rm A} \approx \frac{1}{16\pi^2}\,\alpha_\gamma f\,\frac{M_{\rm dh}^{\rm A} M_{\rm x} c^2}{m_{\rm x} r_{\odot a}^2}\,\frac{\langle\sigma v\rangle}{R_{\rm x}^3}\left[\left(\frac{R_{\rm x}}{r_{\rm c}}\right)^{2\alpha-3}-1\right]. \quad (146)$$

The estimates show [54] that the galactic NeS do not make an appreciable contribution to the total radiation from Andromeda (146).

A similar estimate of gamma radiation flux can also be obtained for the LMC. The character of rotation curves in the case of the LMC indicates that it has lost the greater part of dark-matter halo owing to the interaction with our Galaxy. Hence, as an estimate one may assume the LMC dark-matter mass to be of the same order of magnitude as the baryonic mass:

$$M_{\rm d}^{\rm LMC} \leqslant M_{\rm b}^{\rm LMC}\,.$$

Under these conditions, the distribution (129), (144) does not depend on the dark matter of the LMC. Thus, the LMC must be observed as a distributed source of gamma-radiation and must be defined by relation (145) replacing $r_{\odot a}$ by the corresponding LMC parameters.

The gamma-radiation of other nearby galaxies is described by relations (145), (146) used in the case of Andromeda. Remote galaxies are point sources, and their radiation is given by formula (146).

### 7.6 Comparison of the theory of gamma-radiation from neutralino stars with observations
**7.6.1 Diffusive gamma-radiation.** When comparing theory with observations, it is natural to assume that NeS are just NO that are observed in microlensing. Hence, the following parameters will be used (see Sections 6.2, 2.4):

$$M_{\rm x} = 0.5 M_\odot\,, \qquad R_{\rm x} \approx 4\times 10^{14}\,{\rm cm}\,,$$
$$m_{\rm x} = 10\,{\rm GeV}\,, \qquad H \approx 70\,{\rm km\,s^{-1}\,Mpc^{-1}}\,. \quad (147)$$

The process of neutralino annihilation can be described by a simple expression [19]

$$\langle\sigma v\rangle = \langle\sigma v\rangle_0 \times \frac{r_{\rm g}}{R_{\rm x}} \times \left(\frac{m_{\rm x}}{10\,{\rm GeV}}\right)^2\,,$$
$$\langle\sigma v\rangle_0 \approx (10^{-26}-10^{-27})\,{\rm cm}^3\,{\rm s}^{-1}\,, \quad (148)$$

where $r_{\rm g} = 2GM_{\rm x}c^{-2}$ is the gravitational radius of a NeS.

Proceeding from relations (127), (126), and (147), we shall define the diffusive extragalactic radiation [54]:

$$I_{1\gamma} = 1.4\times 10^{-4}\alpha_\gamma\Omega\left(\frac{f}{0.5}\right)\left(\frac{t_0}{2\times 10^{17}\,{\rm s}}\right)\left(\frac{m_{\rm x}}{10\,{\rm GeV}}\right)$$
$$\times \left(\frac{\rho_{\rm c}}{10^{-29}\,{\rm g\,cm^{-3}}}\right)\left(\frac{4\times 10^{14}\,{\rm cm}}{R_{\rm x}}\right)^4$$
$$\times \left(\frac{M_{\rm x}}{10^{33}\,{\rm g}}\right)^2\left(\frac{\langle\sigma v\rangle_0}{10^{-27}}\right)\frac{\rm photons}{\rm cm^2\,s\,sr}\,. \quad (149)$$

As to the diffusive gamma-radiation from the halo of our Galaxy, we shall note that it depends on the total dark-matter mass $M_{\rm dh}$ and the halo size $R_{\rm h}$. In calculations, we assume as usual [6],

$$M_{\rm dh} = 2\times 10^{12} M_\odot\,, \qquad R_{\rm h} = 200\,{\rm kpc}\,, \qquad r_\odot = 8.5\,{\rm kpc}\,. \quad (150)$$

For the ratio of components we obtain

$$p_\gamma = \frac{I_{2\gamma}^0}{I_{1\gamma}} = \frac{0.36}{\Omega}\left(\frac{f_{\rm h}}{f}\right)\left(\frac{M_{\rm dh}}{2\times 10^{12}M_\odot}\right)\left(\frac{200\,{\rm kpc}}{R_{\rm h}}\right)^{2-\alpha}$$
$$\times \left(\frac{10^{-29}}{\rho_{\rm c}}\right)\left(\frac{2\times 10^{17}\,{\rm s}}{t_0}\right). \quad (151)$$

For the minimum diffuse flux from the halo we have

$$I_{2\gamma}^0 = 6.1\times 10^{-5}\alpha_\gamma\left(\frac{f_{\rm h}}{0.5}\right)\left(\frac{M_{\rm dh}}{2\times 10^{12}M_\odot}\right)\left(\frac{200\,{\rm kpc}}{R_{\rm h}}\right)^{2-\alpha}$$
$$\times \left(\frac{m_{\rm x}}{10\,{\rm GeV}}\right)\left(\frac{4\times 10^{14}\,{\rm cm}}{R_{\rm x}}\right)^4$$
$$\times \left(\frac{M_{\rm x}}{10^{33}\,{\rm g}}\right)^2\left(\frac{\langle\sigma v\rangle_0}{10^{-27}}\right)\frac{\rm photons}{\rm cm^2\,s\,sr}\,. \quad (152)$$

Since in the direction perpendicular to the galactic plane — the direction of the pole — the diffuse flux from the Galaxy is certainly small, the gamma-background $I_{\rm b\gamma}$, i.e., the minimum flux observed in the vicinity of the Sun, is specified by

the expression

$$I_{b\gamma} = I_{1\gamma} + \frac{\pi}{2} I_{2\gamma}^0 = I_{1\gamma}\left(1 + \frac{\pi}{2} p_\gamma\right). \quad (153)$$

Here the factor $\pi/2$ allows for the difference between the gamma-radiation of the halo in the directions towards the pole and towards the anticenter (133). The total flux of diffusive gamma-radiation in each direction $\theta, \phi$ is

$$I_\gamma = I_{1\gamma} + I_{2\gamma}^0 F_\theta + I_{2\gamma}^G(\theta, \phi), \quad (154)$$

where the function $F(\theta)$ is defined by formulae (132) and (133) and the flux $I_{2\gamma}^G$ by relation (139).

Let us now compare the results of the theory with observations. For neutralino annihilation it is natural to consider the gamma-ray fluxes due to $\pi^0$-meson decay [55], i.e., fluxes with energies $E > 100$ MeV. For these energies, the observed gamma-background is, according to Refs [56, 57],

$$I_{b\gamma} \approx 1.5 \times 10^{-5} \frac{\text{photons}}{\text{cm}^2 \, \text{s} \, \text{sr}}. \quad (155)$$

This value agrees with the theory (153), (149), (152) provided that

$$q = \alpha_\gamma \left(\frac{f}{0.5}\right) \Omega \left(\frac{m_x}{10\,\text{GeV}}\right) \left(\frac{4 \times 10^{14}\,\text{cm}}{R_x}\right)^4$$
$$\times \left(\frac{M_x}{10^{33}\,\text{g}}\right)^2 \left(\frac{\langle \sigma v \rangle_0}{10^{-27}}\right) \left(\frac{t_0}{2 \times 10^{17}\,\text{s}}\right) \approx 0.1. \quad (156)$$

Here the coefficient $\alpha_\gamma$ describes the number of photons with energy $E > 100$ MeV generated by the annihilation of 1 GeV of neutralino energy. For the same parameters $q = 0.1$ and $p_\gamma = 0.25$, the latitudinal and longitudinal dependences of the total flux $I_\gamma$ (154) are presented in Figs 18 and 19. One can see that the theory is in agreement with observations.

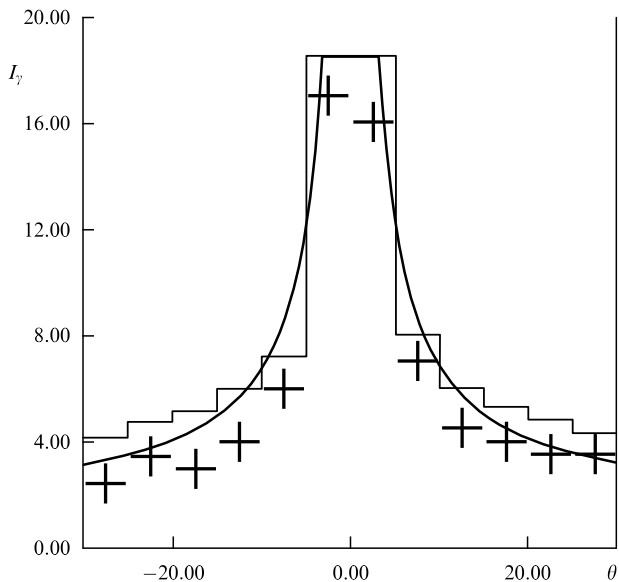

**Figure 18.** Latitudinal dependence $I_\gamma(\theta)$ for $p_\gamma = 0.25$, $q = 0.1$. The solid curve is the theory (154). The data of observations are borrowed from Ref. [57].

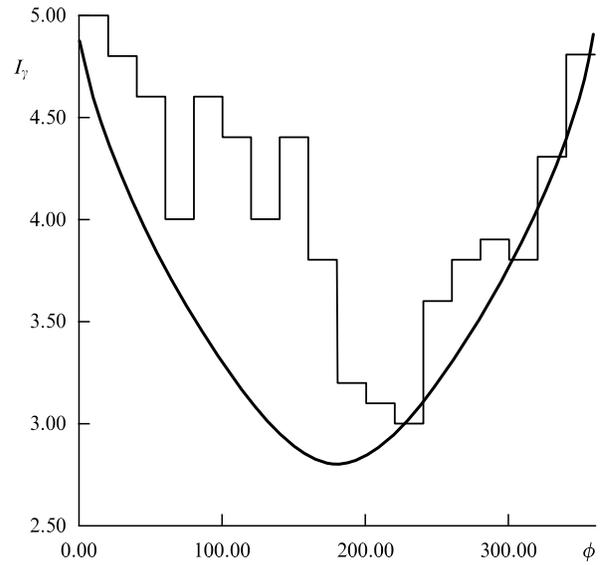

**Figure 19.** Longitudinal dependence $I_\gamma(\phi)$ for $p_\gamma = 0.25$, $q = 0.1$.

It is of importance to emphasize that formula (148) for $\langle \sigma v \rangle$ is obtained under the assumption that neutralino annihilation has a $p$-wave character [58, 59]. Usually (see Ref. [53]), neutralinos with an $s$-wave annihilation channel are considered. Such a process leads in our case to much greater (two-three orders of magnitude) $\langle \sigma v \rangle$ values. Under these conditions, the gamma-ray flux appears to be exceedingly large. For this reason, such particles cannot be treated as candidates for NeS [19, 54, 55].

So, we see that comparison of the theory with observations of diffusive gamma-radiation gives an appreciable limitation on the type of neutralinos in NeS: the $p$-wave process must dominate in the annihilation cross-section. Particles possessing such properties, namely, light photons, have recently been analysed in papers [58, 59]. If the dark matter consists of this type of particle, the NeS theory is in agreement with the diffusive-flux observations.

We shall emphasize that our brief comparison with observations of both galactic and metagalactic components of diffusive gamma-radiation in no way contradicts traditional explanations of the generation of this emission by cosmic rays [57] and active galactic nuclei [60, 61]. The estimates presented here only allowed us to impose a restriction (156) on the properties of annihilating particles and demonstrate a qualitative agreement between the model proposed and the data of observations. If it turns out that neutralino stars do exist, the question of the relative contribution of various radiation generation mechanisms will undoubtedly require a thorough quantitative analysis which is beyond the scope of the present paper.

**7.6.2 Discrete sources of gamma-radiation.** Discrete sources of gamma-radiation with energies $E \geqslant 100$ MeV are partially identified with galactic nuclei, pulsars, and other active objects. There exist, however, a large number of unidentified sources. They were first discovered with the satellite COS B [62]. These sources are now investigated most intensively by the EGRET telescope of the COMPTON observatory.

A new extended version of the EGRET catalogue containing 95 unidentified sources has recently been published [63]. The observed sources are divided into three groups according to their morphology: *em* — extensive or multiple sources, *C* — doubtful and extended sources, and *P* — definite point sources within the scope of the EGRET diagram (1°).

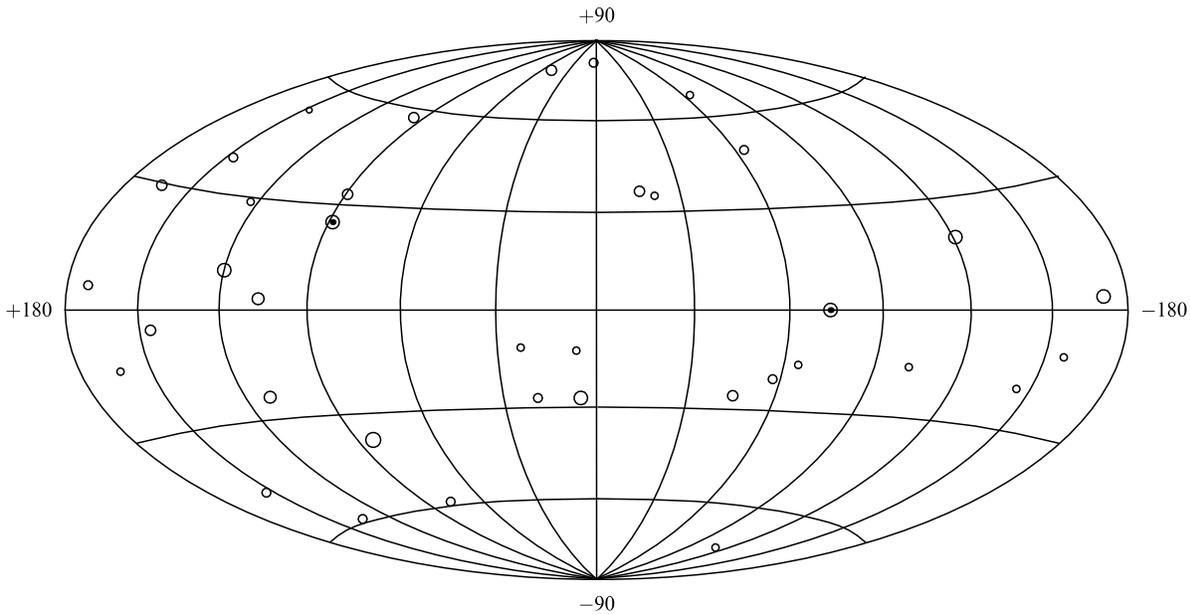

**Figure 20.** Position of *P*-sources in the sky in galactic coordinates *l*, *b* according to EGRET [64].

Naturally, it is only the last group that may be treated as possible candidates for sources of observed NeS radiation. In the extended version of EGRET catalogue, this group now amounts to 40 sources [64]. The position of *P*-sources in the sky in the galactic coordinates *l* and *b* is indicated in Fig. 20. As can be seen from the figure, their distribution across the sky is sufficiently uniform and small asymmetries may be associated with the non-equilibrium distribution of the background radiation and the time of observation (see Fig. 1 in Ref. [63]).

Figure 21 presents a $\log N - \log S$ diagram for these sources. It also well confirms the uniform and isotropic distribution of sources: all deviations from the $-3/2$ law lie within experimental error. The isotropy and the $\log N - \log S$ curve are in full agreement with the fact that the *P*-sources of gamma-radiation (or the greater part of them) are a direct observation of NeS. By the $\log N - \log S$ curve one can determine the intensity of the sources in the vicinity of the Sun at a characteristic distance (142). This intensity is

$$I_0 \approx 5 \times 10^{-7} \frac{\text{photons}}{\text{cm}^2\,\text{s}}\,. \tag{157}$$

Allowing for the value of the parameter $\alpha_\gamma m_x \langle \sigma v \rangle_0$ from Eqn (156), one can find the value of the same flux $I_0$ predicted by the theory from formula (143):

$$I_0 = \frac{5 \times 10^{-7}}{\Omega} \left(\frac{f}{0.5}\right)^{-1/3} \left(\frac{2 \times 10^{17}\,\text{s}}{t_0}\right) \left(\frac{M_x}{10^{33}\,\text{g}}\right)^{1/3}$$
$$\times \left(\frac{M_{\text{dh}}}{2 \times 10^{12} M_\odot}\right)^{2/3} \left(\frac{200\,\text{kpc}}{R_\text{h}}\right)^{0.8} \left(\frac{p}{50}\right) \frac{\text{photons}}{\text{cm}^2\,\text{s}}\,. \tag{158}$$

This implies that the flux predicted by the theory agrees with the observed value (157) under the assumption that $p \sim 50$, i.e., that the galactic NeS on average have a significant baryonic core (138) of mass $M_\text{b} \geqslant 0.3 M_x$. It should be stressed that the expression for the discrete flux (158) also agrees with the observed diffuse flux (155) because in its derivation we used relation (156).

Also worthy of attention is another assumption according to which unidentified *P*-sources are active galactic nuclei. Testifying in favor of this hypothesis are the observed gamma-radiation spectra [61], as well as a possible flux instability. However, the identified sources of high-energy gamma-radiation have always been identified with blasar type active galactic nuclei that show powerful radio emission. In our case there is no such emission, and therefore one can only speak of a new type of object.

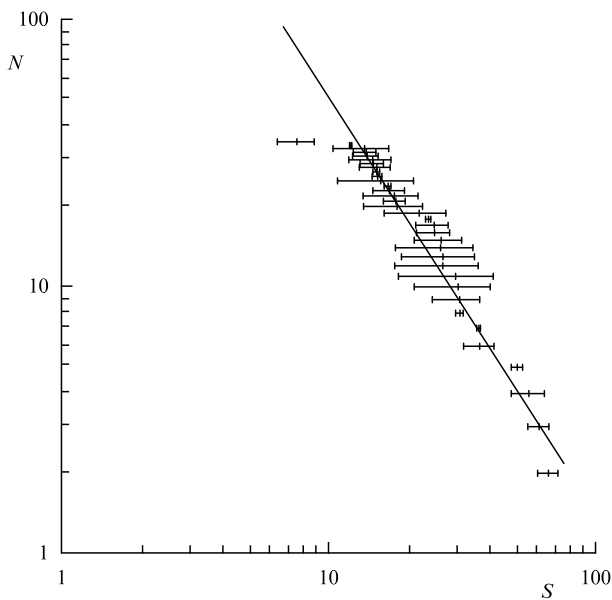

**Figure 21.** $\log N - \log S$ diagram for *P*-sources according to EGRET [64].

### 7.6.3 Distributed gamma-radiation from galaxy M31.
In determining the gamma-radiation from the Andromeda nebula we shall employ the following parameters

$$M_{\rm dh}^{\rm A} = 3 \times 10^{12} M_\odot, \qquad R_{\rm ha} = 250 \, {\rm kpc}, \qquad r_{\odot {\rm a}} = 600 \, {\rm kpc}.$$

Then, according to Eqn (145), the flux from the Andromeda nebula is

$$I_\gamma^{\rm A} \approx 1.1 \times 10^{-5} K(\theta) \alpha_\gamma \left(\frac{f}{0.5}\right) \left(\frac{M_{\rm dh}^{\rm A}}{3 \times 10^{12} M_\odot}\right)$$
$$\times \left(\frac{M_{\rm x}}{10^{33} \, {\rm g}}\right)^2 \left(\frac{4 \times 10^{14} \, {\rm cm}}{R_{\rm x}}\right)^4 \left(\frac{250 \, {\rm kpc}}{R_{\rm ha}}\right)^2$$
$$\times \left(\frac{m_{\rm x}}{10 \, {\rm GeV}}\right) \left(\frac{\langle \sigma v \rangle_0}{10^{-27}}\right) \frac{\rm photons}{\rm cm^2 \, s \, sr},$$

$$K(\theta) = \frac{1}{|\sin \theta|^{\alpha - 1}} \arctan \sqrt{\left(\frac{0.416}{\sin \theta}\right)^2 - 1}.$$

This means that EGRET (with its resolving ability of 1°) can receive a flux $I_\gamma$ from the center of Andromeda of the order of

$$I_{0\gamma}^{\rm A} \approx 1.3 \times 10^{-8} \frac{\rm photons}{\rm cm^2 \, s}.$$

Here we have again employed relation (156). As follows from Eqn (145), when moving from the galactic center, the gamma-ray flux in the halo ($\theta > 1°$) decreases

$$I_\gamma^{\rm A}(\theta) = I_{0\gamma}^{\rm A} \times \left(\frac{1°}{\theta°}\right)^{0.8}$$

and vanishes completely at $\theta \approx 20°$. According to Eqns (146) and (156), the total integral flux from the halo of Andromeda is equal to

$$I_\gamma^{\rm A} \approx 2 \times 10^{-7} \frac{\rm photons}{\rm s}.$$

Such a flux might, in principle, be recorded by EGRET. However, it is created by an extensive source with an angular scale of the order of 20°. The background radiation is much higher on this scale, and thus the question of the possibility of signal detection remains open.

### 7.7 Non-compact objects as sources of gamma-ray bursts
One of the most intriguing astrophysical problems of the last decades is the problem of gamma-ray bursts. In spite of the considerable joint efforts of observers and theoreticians, nothing is known today of how gamma-ray bursts are generated and even where they come from. Two basic models — the cosmological and the giant halo — are now under discussion [65]. According to the cosmological model, the sources of gamma-ray bursts are at cosmological distances from one another; the giant halo model suggests that the sources are located in the halo of the Galaxy.

The model of a giant dark-matter halo (GDMH) was proposed in Ref. [66]. This model considers relic neutron stars as sources of gamma-ray bursts. These stars must have the same density distribution as the dark matter in the Galactic halo, and this fact gave the name to the model. According to the parameter values (150), the size of the halo is about $R_{\rm h} \approx 200$ kpc. This model accounts for the basic statistical properties of gamma-ray bursts, namely, their spherical symmetry and a substantial condensation towards the center, which can be expressed in the form of a $\log N - \log S$ curve [67]. Furthermore, the GDMH model well describes the weak mean anisotropy in the distribution of gamma-ray bursts [68] and the local anisotropy [69] observed by the BATSE group at the COMPTON observatory [70] within the first year of measurements. According to this model, the characteristic energy released by a gamma-ray source in one burst is $E_\gamma \sim 10^{41}$ erg.

The weak point of this model is the assumption that sources of gamma-ray bursts are neutron stars. The point is that the analysis of primary metallicity in the Galaxy suggests the limitation $N \leqslant 10^7$ on the total number of such stars. This means that each relic neutron star must yield $10^5 - 10^6$ gamma-ray bursts because the total number during the lifetime of the Universe must be $3 \times 10^{12}$ following from the observed frequency of bursts. Such a large number of gamma-ray bursts produced by a single neutron star is extremely difficult to explain both energetically and from the viewpoint of the mechanism of the process. Similar difficulties are encountered by models that regard high-speed neutron stars as candidates for sources of gamma-ray bursts [71]. These stars gain high speeds from the non-symmetric explosion of supernovae and fly out of the Galaxy. The number of such stars is only of the order of $10^7$ and the same difficulties occur because a multiple repetition of bursts is required. For the model of neutron stars as sources of gamma-ray bursts these difficulties have been discussed in many papers [71, 72].

There are no such difficulties when NO are thought of as possible sources of gamma-ray bursts in the GDMH model [73]. In the first place, the NO distribution coincides with the dark-matter distribution. This leads to the same agreement with the principal observed statistical properties of gamma-ray bursts as in the GDMH model. Moreover, the total number of NO in the galactic halo is ($\sim 10^{12} - 10^{13}$) which approximately corresponds to the expected number of gamma-ray bursts, and thus burst repetition is not required. As a concrete mechanism of gamma-ray bursts, we consider in Ref. [73] the process which can be defined as an explosion of the central baryonic body of a NO.

We shall point out the main parameters of the model [73] and briefly describe the main processes that may lead to an 'explosion'. As shown above (73), under ordinary stationary conditions, the mass of the dark matter trapped in a baryonic body, $M_{\rm n}$, makes up

$$M_{\rm n} \sim 10^{25} - 10^{26} \, {\rm g}.$$

To generate a gamma-ray burst, annihilation of only a fraction of a percent of this mass will suffice. However, the annihilation cross-section (148) is small, so generally the annihilation process proceeds smoothly. The characteristic neutralino annihilation time appears to be of the order of the lifetime $t_0$ of the Universe.

The energy released in smooth annihilation heats the central region of the baryonic body. Its surface temperature $T_{\rm s}$ is determined by the balance between the energy released and the radiation and is $T_{\rm s} \approx 500 - 1000$ K. Thus, the baryonic body is heated and may serve as an IR radiation source [$\lambda \sim (3-8)$ μm] with intensity $P \sim 10^{28} - 10^{29}$ erg s$^{-1}$.

The total gravitational energy of a baryonic body is equal to

$$E = 3 \times 10^{45} \left(\frac{M_{\rm b}}{5 \times 10^{31} \, {\rm g}}\right)^2 \, {\rm erg}.$$

In the case that a baryonic body loses stability, which leads to its 'explosion', this energy (or even a fraction) is absolutely sufficient to produce a gamma-ray burst. In Ref. [73] we propose three basic mechanisms that may cause an 'explosion'.

**1. Superheating of the central part of a baryonic body.** The stationary state of a baryonic body considered above suggests a balance in the entire body that provides the necessary heat flow towards the surface. However, with a sufficiently large quantity of neutralinos, $M_n \geqslant 10^{26}$ g, a great superheating of the central part (especially for large neutralino masses $m_x$) is possible which may ultimately cause a violation of the pressure balance and the destruction of the baryonic body.

**2. Thermonuclear heating.** Neutralino annihilation inside a star induces the generation of high-energy protons and gamma-photons. When interacting with baryons, these energetic particles produce $D_2^+$ and other light elements. As a result, the number of $D_2$ and other particles continuously increases in the center of the baryonic body. According to Ref. [73], a great number could be accumulated within the lifetime of the Universe:

$$N_{D_2} = \gamma_{D_2} \frac{Q_{n0}}{m_x c^2} t_0 \, ;$$

$$N_{D_2} = 1.6 \times 10^{18} \gamma_{D_2} \left(\frac{M_n}{10^{25} \text{ g}}\right)^2 \left(\frac{10^6 \text{ K}}{T^*}\right)^{-1/2}$$

$$\times \left(\frac{m_x}{10 \text{ GeV}}\right)^{-1/2} \left(\frac{\rho_0}{1 \text{ g cm}^{-3}}\right)^2 \left(\frac{\langle \sigma v \rangle_0}{3 \times 10^{-27}}\right) \text{cm}^{-3} \, .$$

Here $\gamma_{D_2}$ is the coefficient of transformation of 1 GeV protons into $D_2^+$ ions. Such a large amount of $D_2$ at a temperature $T_c \sim 10^6$ K in the center may lead to a thermonuclear reaction accompanied by a rapid heating and an explosion of the star.

**3. Self-entrapment of neutralinos.** The neutralino density $\rho_0$ in the central part of the body is normally small compared to the baryon density $\rho_b$. This condition can be violated if the neutralinos start concentrating in the central part of the core

$$r_n \leqslant 3 \times 10^8 \left(\frac{M_n}{10^{26} \text{ g}}\right) \text{cm} \, . \quad (159)$$

This process becomes possible if the neutralino temperature is small, $T^* < 10^3 - 10^4$, or the mass $m_x$ is large so that the density is strongly peaked near the center. If the conditions (159) are met, $\rho_n > \rho_b$ and the neutralinos are trapped by their own gravitational field. This causes the state of equilibrium in the center to change completely, and neutralino self-heating may start as a result of annihilation and scattering on baryons. A remarkable property of self-trapping is that heating strengthens compression. As a result, the process becomes explosive.

The possibility of the indicated processes is discussed in more detail in Ref. [73].

## 8. Conclusion

Observations of microlensing and the theory of small-scale dark-matter structure presented here pose a fundamental question of the origin of the bulk dark matter. If the interpretation [16, 17] is correct, the dark matter would appear to consist mainly of baryonic matter. This would necessitate a revision of the conclusions of the theory of nucleosynthesis, the new questions will arise on the problem of the formation of galaxies, their structure, etc.

If the hypothesis on small-scale non-baryonic dark-matter structure turns out to be valid, this will mean the existence of an essentially new type of structure, i.e., gravitationally bound kinetic formations with characteristic masses less than or of the order of the solar mass and scales of the order of that of the solar system. These objects contain a considerable part or even the bulk of matter in the Universe. They appear before recombination and must have an appreciable effect on the formation of stars and galactic structure.

If it turns out additionally that the non-baryonic dark matter consists of annihilating particles (neutralinos, heavy neutrinos), then the NO, which in this case are called neutralino stars (NeS), must also be observed in the form of powerful sources of gamma-radiation that substantially affect the level of gamma-background and are possibly even the source of the mysterious gamma-ray bursts. On the other hand, observational data impose significant restrictions on the possible form of dark-matter particles. This is why the study of the problems considered here is of crucial interest for cosmology.

In conclusion we note some observational consequences of the theory.

**I. Microlensing**

(1) The difference between theoretical light curves in microlensing by compact and non-compact bodies is not large, but obviously quite accessible for experimental verification (Section 6). In this connection we should specially emphasize the possibility of a notable improvement of the results of observations using telescopes installed from cosmic apparatus. In particular, measurements on the Hubble telescope seem now most promising.

(2) We have considered microlensing on spherical objects. Deviation from spherical symmetry in the case of binary stars is rather significant for NO and may lead to definitely observable effects [74]. Analogous observed deviations from spherical symmetry also occur when objects fly by at a sufficiently close distance.

(3) Strongly elongated wings in the light curves (Section 5.3) imply that NO have a considerably larger lensing cross-section at small amplitudes. It should therefore be expected that they are more strong pronounced in the generation of intensity fluctuations when quasar radiation passes through galactic halos. We stress that microlensing is peculiar in that it provides similar fluctuations in the optical and radio-frequency bands.

(4) It is noteworthy that the discovery of even one NO in the course of microlensing may prove the existence of non-baryonic matter and non-baryonic small-scale structure because baryonic bodies (or gas clouds) of such a mass and size cannot exist (Section 3.5).

**II. Gamma-radiation**

In the discussion in Section 7.6 of the unidentified point sources of high-energy gamma-radiation ($P$-sources) discovered by EGRET, two hypotheses were put forward concerning the nature of these sources: they are either active galactic nuclei of a new yet unknown type or NeS. If a considerable part of the indicated sources turn out to be neutralino stars, then

(1) The theory predicts that the unidentified $P$-sources of gamma-radiation must be distributed isotropically and obey the $-3/2$ law for their $\log N - \log S$ curve.

(2) The theory predicts the existence of a vastly spread ($\sim 20°$) distributed source with quite definite properties around the Andromeda galaxy. The discovery and detailed investigation into the structure of this source would mean the discovery of the fact that the halo consists of NeS.

(3) The theory predicts that owing to the presence of a baryonic core, $P$-sources of gamma-radiation must have as an accompanying component thermal IR radiation in the range $\lambda \sim (3-8)$ μm with a radiation temperature $T \sim 500-1000°$ and intensity $\tilde{I} \sim 10^{28}-10^{29}$ erg s$^{-1}$ [73]. The discovery of thermal IR sources and their identification with gamma-ray sources would mean, in fact, the discovery of NeS and would confirm the presence of a baryonic body.

Other possibilities connected with the direct determination of the distance to $P$-sources by way of observation of their motion on the celestial sphere and observation of microlensing and gamma-radiation for one and the same source are pointed out in Ref. [54].

It should be stressed that comparing the theory with the gamma-radiation observational data allows already to obtain some important conclusions concerning the origin of the dark-matter particles. The relation (156) connecting the mass $m_x$ of the particles, the annihilation cross-section (148), and the coefficient of gamma-radiation generation is not performed in the nowaday standard model of neutralino [53].

Therefore, if further investigations will prove that microlensing is accomplished by the non-compact dark-matter objects, then either special modifications of the standard model (like, e.g., in Refs [58, 59]) would be needed or the dark matter consists of some other kind of particles.

Further theoretical and experimental investigation into the problem considered here is of fundamental interest.

The authors are grateful to V L Ginzburg for fruitful discussions and permanent interest in the work, and also to W I Aksford and M V Sazhin for discussions and comments.

This work was partially supported by the Russian Foundation for Basic Research (grant 96-02-18217). One of the authors (V.S.) is indebted to the Landau Foundation (Julich, Germany) and International Soros Foundation (ISSEP) for sponsoring.